\documentclass[12pt]{emulateapj}
\def \bea {\begin{eqnarray}}
\def \ena {\end{eqnarray}}
\def \bee {\begin{equation}}
\def \ene {\end{equation}}
\def    \bJ     {\bf J}
\def    \ba     {\bf a}

\def    \be     {\bf e}
\def    \bk     {\bf k}

\def    \bQ     {\bf Q}
\def    \bB     {\bf B}

\def \me {\hat{\bf e}}

\def    \bG     {{\bf \Gamma}}
\def \mc {\mbox{cos}}
\def \ms {\mbox{sin}}

\begin{document}
\shorttitle{Radiative torques alignment}
\shortauthors{Hoang \& Lazarian}
\title{Grain alignment induced by radiative torques: effects of internal relaxation of energy and complex radiation fields}

\author{Thiem Hoang  \& A. Lazarian}
\affil{Dept. of Astronomy, University of Wisconsin, Madison, WI53706; hoang; lazarian@astro.wis.du}

\begin{abstract}
Earlier studies of grain alignment dealt mostly with interstellar grains that have strong internal relaxation of energy which aligns grain axis of maximum moment of inertia with respect to rain's angular momentum. In this paper, we study the alignment by radiative torques for large irregular grains, e.g., grains in accretion disks, for which internal relaxation is subdominant. We use both numerical calculations and the analytical model of a helical grain introduced by us earlier. We demonstrate that grains in such a regime exhibit more complex dynamics. In particular, if initially the grain axis of maximum moment of inertia makes a small angle with angular momentum, then radiative torques can align the grain axis of maximum moment of inertia with angular momentum, and both axis of maximum moment of inertia and angular momentum are aligned with the magnetic field when attractors with high angular momentum (high-J attractors) are available. For the alignment without high-J attractors, beside the earlier studied attractors with low angular momentum (low-J attractors), there appears new low-J attractors. The former and later cases correspond to the alignment with long axes perpendicular and parallel to the angular momentum, respectively.
In addition, we also study the alignment of grains in the presence of strong internal relaxation, but induced not by a radiation beam as in earlier studies, instead, induced by a complex radiation field, that can be decomposed into dipole and quadrupole components. We found that in this situation, the parameter space $q^{max}$, for the existence of high-$J$ attractors in trajectory maps is more extended, which entails higher degrees of polarization expected. Our obtained results are useful for modeling polarization arising from aligned grains in molecular clouds and accretion disks.
\end{abstract}

\section{Introduction}

Magnetic fields play a crucial role in many astrophysical processes (e.g., star formation, accretion disks, cosmic ray transport). An important and easily available way to map magnetic fields is to observe the polarization of emission or absorption by aligned grains (Goodman et al. 1995; Hildebrand et al. 2000, 2002; Crutcher et al. 2004). The reliable interpretation of polarimetry in terms of magnetic fields requires, however, a solid understanding of the alignment of grains. This is the major motivation of work in the field of grain alignment. An additional motivation to understand when grains get aligned comes from the necessity of separating the polarization by dust grains from the polarized Cosmic Microwave Background (CMB) signal (see Lazarian 2008). 

Grain alignment is a complex area of studies related to many physical effects of different time scales (see review by 
Lazarian 2007). It is also the research area associated with the names of Lyman Spitzer and Ed Purcell, who decisively contributed to understanding of a number of fundamental physical processes of grain alignment. 

A number of mechanism has been known to induce grain alignment (see Lazarian 2007). However, recently
grain helicity has been discussed as the necessary element of successful alignment. Note that the idea of grains having twist and therefore
interacting differently with left and right polarized photons can be traced back to the pioneering work by
Dolginov \& Mytrophanov (1976). A more recent work in the area showed that a similar effect is present when
irregular grains interact with a gaseous flow (Lazarian \& Hoang 2007b).\footnote{Such flows arise naturally in the reference of a grain as the grain interacts with ubiquitous MHD turbulence (Lazarian \& Yan 2002; Yan \& Lazarian 2003; Yan, Lazarian, \& Draine 2004).}

In this paper we concentrate on studying grain alignment by radiative torques (hereafter RATs). However, the intrinsic similarity of the radiative and mechanical torques acting on helical grains (see discussion in Lazarian \& Hoang 2007b) makes our present results applicable to the alignment by mechanical torques. Note that for the alignment of helical grains, the magnetic field acts only through inducing the Larmor precession. Therefore the alignment may potentially occur with both longer and shorter grain axes perpendicular to the magnetic field. Observations of interstellar dust indicate that the alignment with the long axes perpendicular to the magnetic field is 
dominant in the diffuse gas. Therefore we shall call this alignment {\it right alignment}, and the opposite type of alignment we shall call {\it wrong alignment}. Different alignment processes produce either right or wrong alignment and the goal of the grain alignment theory is to predict correctly both the direction and the degree of the expected alignment.

 As dust grain scatters or absorbs photons it experiences radiative torques. These torques can be stochastic or regular. Stochastic RATs arise from, for instance, a spheroidal grain randomly emitting or absorbing photons. The latter process, for instance, was invoked by Harwit (1970) in his model of grain alignment based on grains being preferentially spun up in the direction perpendicular to the photon beam. However, Purcell \& Spitzer (1971) showed that the randomization arising from the same grain emitting thermal photons makes the achievable degree of grain alignment negligible. In fact, they showed that stochastic RATs arising from grain thermal emission is an important process of grain randomization. More recently, grain thermal emission was analyzed as a source of the excitation of grain rotation as well as its damping in relation to the rotation of tiny spinning grains that are likely to be responsible for the so-called anomalous foreground emission (Draine \& Lazarian 1998). 

RATs were first discussed by Dolginov (1972) in terms of chiral, e.g. hypothetical quartz grains. The idea is that starlight passing through such grains would spin them up. Later, Dolginov \& Mytrophanov (1976) considered an irregular grain model of two spheroids twisted to each other, made of more accepted materials, e.g. silicate grains, and claimed that these grains will be both spun up and aligned by {\it regular RATs}. This work was, unfortunately, mostly ignored for 20 years. 

Grain alignment by RATs drew much attention when Bruce Draine modified the Discrete Dipole Approximation Code(Draine \& Flatau 2004, hereafter DDSCAT) and enabled later researchers to calculate RATs for irregular grains. \footnote{The only work on RATs before that was Lazarian (1995) which corrected some of the points in the Dolginov \& Mytrophanov (1976) study, but underestimated the importance of RATs.}

Draine \& Weingartner (1996, hereafter DW96) treated RATs as a kind of pinwheel torque, and they found that RATs can accelerate grains to suprathermal rotations. Later, Draine \& Weingartner (1997, hereafter DW97) introduced many essential elements of the modern treatment of RATs, for instance, phase trajectory maps. They confirmed Dolginov \& Mytraphanov (1976) hypothesis that RATs can produce the alignment by their own. 
However, the important moments of grain dynamics, so-called crossovers, were not treated correctly in the paper.
Crossovers, are special events in the dynamics of grains rotating subject to regular torques. During crossovers grain angular velocity
approaches its minimal value and models of crossovers caused by pinwheel torques in Spitzer \& McGlynn (1979) and in Lazarian \& Draine (1997) envisaged grain flipping with the direction of the angular momentum staying the same. DW97, however, assumed that the RAT crossovers are different and during the RAT-induced crossovers angular momentum changes its direction to the opposite. This treatment of crossovers resulted in grains having cyclic trajectories, which were, in fact, an artifact of the adopted 
treatment, as was shown later (Lazarian \& Hoang 2007a, henceforth LH07a). In fact, LH07a showed that instead of cyclic trajectories, one gets situations when RATs tend to {\it slow} grains down.

A more general analysis for dynamics of RAT alignment was done in Weingartner \& Draine (2003, henceforth WD03), where thermal fluctuations within grain were accounted for. These fluctuations were shown by Lazarian (1994, see also Lazarian \& Roberge 1997) to induce grain thermal wobbling, which amplitude gets larger as the grain angular momentum approaches its thermal values.  For their choice of parameters, WD03 found that an attractor with low angular momentum, so-called, low-J attractor, appears on the trajectory phase maps.\footnote{Conventionally, an attractor with angular momentum $J$ larger than the thermal value is called a high-J attractor, and the attractor with $J$ of the order of thermal value is called low-J attractor.} The study, however, was limited in terms of parameter space explored. In fact, it was a numerical study of alignment of a single grain subjected to the electromagnetic wave of a single frequency at a single incident direction. Therefore implications of the study for the problem of the RAT alignment were difficult to evaluate.

We feel that the numerical work above done with DDSCAT should be treated as an experimental attack on the problem of the RAT alignment. The limitations of such an approach are self-evident. For instance, to obtain predictions one should analyze many hundreds of different grain shapes and many wavelengths and many incident directions. The study
in DW97 included 3 shapes and resulting torques looked very different, which made one wonder what causes the alignment\footnote{In relation to this one of us (AL) recalls that Lyman Spitzer after studying the work on RATs suggested that one should seek for some simple trigonometric fits to torques that would produce the alignment.}. At the same time, unfortunately, the analytical study in Dolginov \& Mytrophanov (1976) was in error.

The first analytical model (henceforth AMO) of RATs consistent with the numerical calculations was proposed in LH07a. This study provided simple analytical expressions for the RATs (see also \S 2). In particular, the functional forms of the torques acting on grains of different shapes are similar, with the difference between the grains of different shapes amounting to a single parameter, termed $q^{max}$-ratio, which is the ratio of the magnitudes of two first components of torques in the grain symmetry system of reference where the radiation direction is the symmetry axis. AMO allowed theoretical predictions and the analytical treatment of the RAT alignment. As a result, studies of a parameter space seized to be an insurmountable  problem. In LH07a, similar to DW97, the thermal fluctuations were intentionally disregarded and in such a setup the low-$J$ attractors, rather than cyclic trajectories, were reported. In fact, it became clear that RATs, over a large part of the parameter space, {\it spin down} rather than {\it spin up} grains. Using AMO, LH07a obtained the values $q^{max}$ for which grains have only low-$J$ attractors and when they have both low-$J$ and high-$J$ attractors simultaneously.

AMO was extensively used in all our papers that followed. For instance, in Hoang \& Lazarian (2008a, henceforth HL08a), we studied the RAT alignment in the presence of thermal fluctuations using both AMO and DDSCAT for an extended sample of grain shapes, radiation direction and wavelength. We found that {\it irregular} grains do not stop completely, but rotate at a rate, which is comparable with the rate of thermal rotation at the dust temperature, which agrees with an earlier example in WD03. Importantly, HL08a considered the RAT alignment in the presence of gaseous bombardment and reported a new effect, namely, the transfer between the low-$J$ and high-$J$ attractors, in situations when the both low-$J$ and high-$J$ attractors were present simultaneously\footnote{Note that the parameter space for the existence of low-$J$ and high-$J$ attractors are similar and known through the LH07a study.}. Thus, counterintuitively, the collisions were found to {\it increase} the alignment. 

The most important practical implication of the LH07a and HL08a studies was that, over a large range of the parameter space of $q^{max}$-ratio and angles between the magnetic field and radiation direction, the grains rotate thermally and therefore, as a result the aforementioned thermal fluctuations of grains, and exhibit
the reduced alignment of grain axes with respect to ${\bf J}$. This opened possibilities of estimating the expected alignment and comparing it to the alignment inferred from observations. Importantly enough, Lazarian \& Hoang (2008, henceforth LH08) reported, that in the presence of superparamagnetic inclusions (Jones \& Spitzer 1967, Mathis 1986, Bradley 1994, Martin 1994, Goodman \& Whittet 1994), the high-$J$ attractors {\it always} exist. This finding entails a very non-trivial effect, namely, that,
in the presence of both superparamagnetic inclusions, grains, that otherwise would rotate subthermally, get into the state of fast suprathermal rotation. According to HL08a, this also means that {\it all} grains eventually get into the state of fast rotation and {\it perfect} alignment. From the point of view of observations, this opens prospects of testing the existence of superparamagnetic inclusions by measuring the degree of polarization.

This paper continues our studies in LH07a and HL08a. The major issues we address below are (1) the effect of internal relaxation, and (2) the effect of complex radiation field on the grain alignment.

The first issue is important in terms of the alignment of large grains, e.g. grains within proto-planetary accretion disks. The alignment of such grains was assumed in a recent modeling of polarization from T-Tauri accretion disks in Cho \& Lazarian (2007). Grains which are important for such a modeling may, however, should be sufficiently large that the internal relaxation time for them is longer than the alignment time. Therefore, how grains get aligned in the absence of efficient internal relaxation is an important question.

The second issue is vital for modeling the polarization from both disks and dark clouds. The latter modeling which used the RAT alignment was undertaken in Cho \& Lazarian (2005), Pelkonen et al. (2007) and Bethell et al. (2007). All these studies assumed rather naive models of alignment. In particular, the actual modeling would require taking into account the actual complex radiation field that is  being experienced by the grain, rather than to assume that the radiation is coming from a point source.

In what follows, in \S 2  we briefly describe the AMO, grain torque-free motion, and calculations of RATs. In \S 3 we compare the Barnett and nuclear relaxation times with the radiative alignment time and identify grain size when the internal relaxation can be disregarded. In \S 4 we study the properties of RATs and the resulting RAT alignment for large grains in the absence of the internal relaxation. Grain alignment by RATs arising from dipole and quadrupole components of radiation fields is studied in \S 5. We discuss our findings in \S 6, and our summary is presented in \S 7.

\section{Radiative torques}
\subsection{Definition of radiative torques}
\begin{figure}
\includegraphics[width=0.49\textwidth]{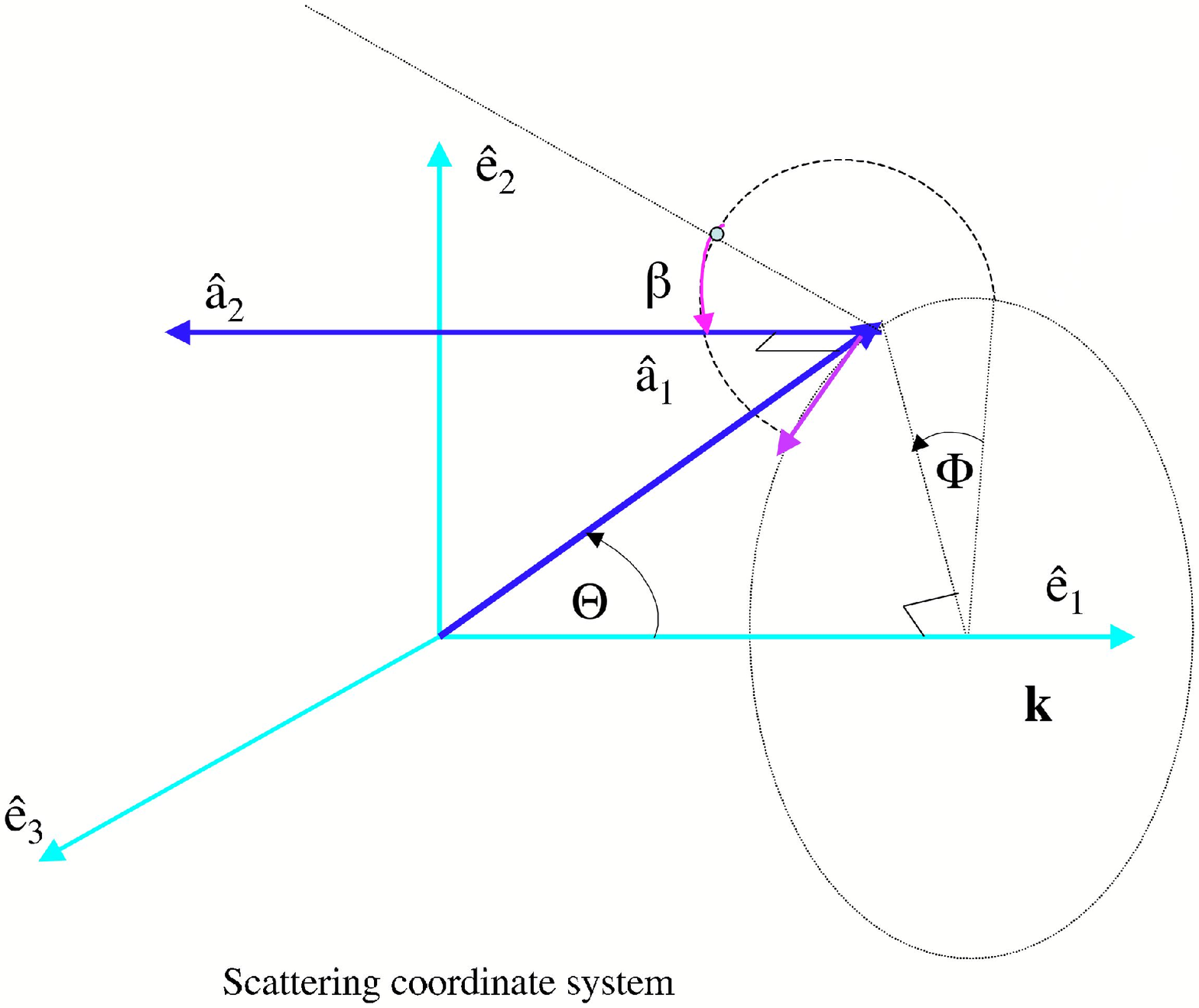}
\includegraphics[width=0.49\textwidth]{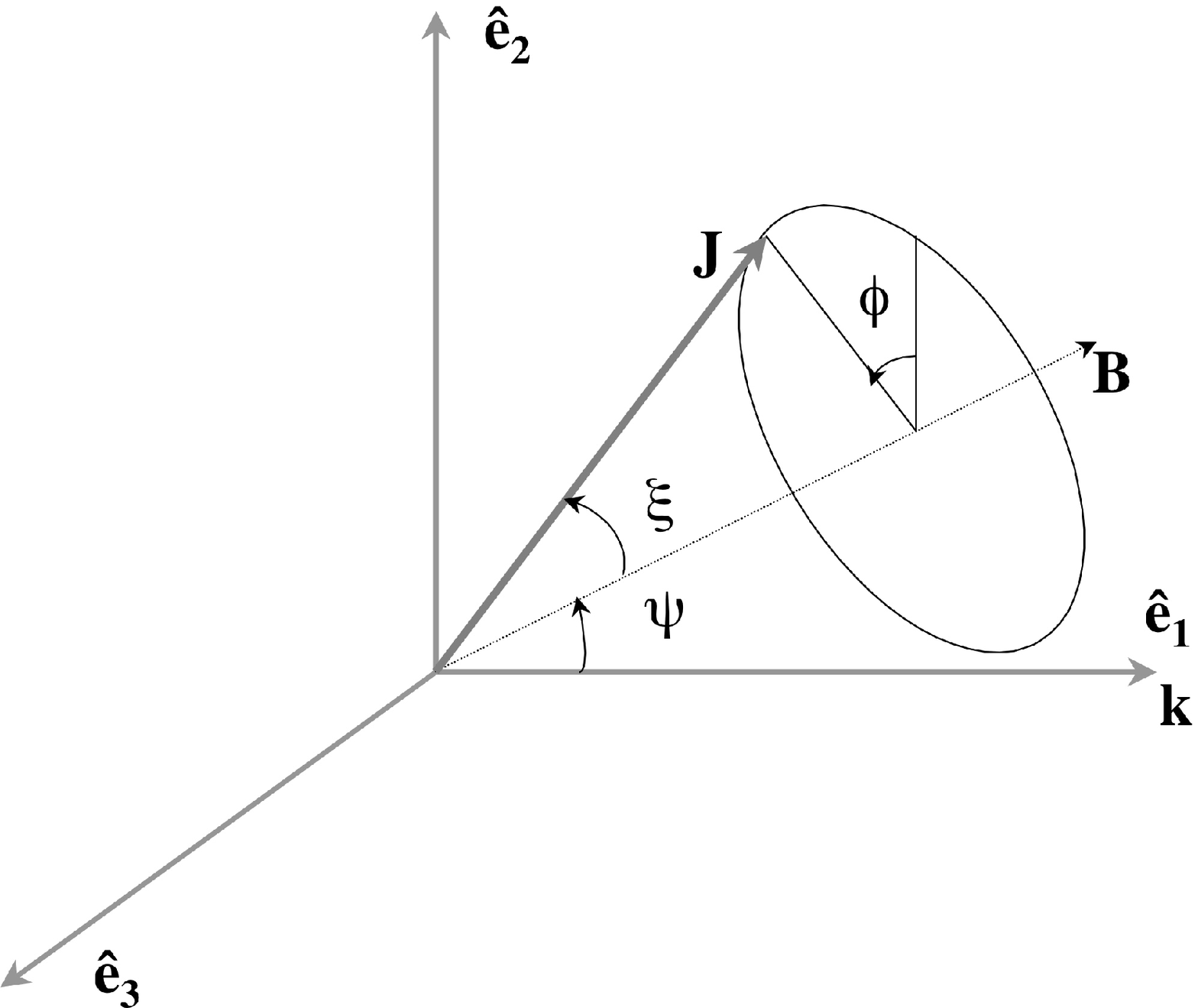}
\caption{{\it Upper panel}: Orientation of a grain, described by three principal axes $\hat{a}_{1},\hat{a}_{2}$ and $\hat{a}_{3}$, in the lab coordinate system  $\hat{e}_{1},\hat{e}_{2}, \hat{e}_{3}$ is defined by three angles $\Theta, \beta$ and $\Phi$. The direction of incident radiation beam ${\bf k}$ is along $\hat{e}_{1}$. {\it Lower panel: }The orientation of angular momentum $\bJ$ in the lab coordinate system, $\psi$ is the angle between the magnetic field ${\bf B}$ and ${\bf k}$, $\xi$ is the angle between ${\bf J}$ and ${\bf B}$, and $\phi $ is the Larmor precession angle of ${\bf J}$ about ${\bf B}$.}
\label{f2}
\end{figure}

Considering only the anisotropic component of the radiation field, the radiative torque resulting from the interaction of a radiation beam with a grain is then defined by
\bea 
\bG_{rad}=\frac{{u}_{rad}a^{2}\bar{\lambda}}{2}\gamma{\overline{\bQ}}_{\Gamma}(\Theta, \beta, \Phi),\label{eq7}
\ena 
where $a$ is the grain size, $\gamma$ is the degree of anisotropy, $\bar{\lambda}$ and $
{u}_{rad}$ are the mean wave length and total energy density of the radiation
field, respectively; ${\bf Q}_{\Gamma}$ is the RAT efficiency vector depending on angles $\Theta, \beta$ and $\Phi$, and its overline denotes averaging over the spectrum of radiation field (DW96); $\Theta$ is the angle between the axis of maximum moment of inertia ${\bf a}_{1}$ and the radiation direction ${\bf k}$, $\beta$ is the rotation angle about $\ba_{1}$ and $\Phi$ is the precession angle of $\ba_{1}$ about $\bk$ (see Fig. \ref{f2}{\it upper}).  In the lab coordinate system $\me_{1}\me_{2}\me_{3}$, ${\bf Q}_{\Gamma}$ is decomposed as
\bea
{\bf Q}_{\Gamma}(\Theta, \beta, \Phi)&=&Q_{e_{1}}(\Theta, \beta, \Phi)\me_{1}+Q_{e_{2}}(\Theta, \beta, \Phi)\me_{2}\nonumber\\
&&+Q_{e_{3}}(\Theta, \beta, \Phi)\me_{3}.\label{eq7}
\ena

\subsection{Analytical model of RATs}

LH07a proposed an analytical model of RATs arising from the interaction of a helical grain, consisting of a mirror attached to an ellipsoid body with a radiation beam. Although this model was derived in the geometric optics approximation, LH07a proved that it represents well RATs obtained by DDSCAT.

 \begin{figure}
\includegraphics[width=0.5\textwidth]{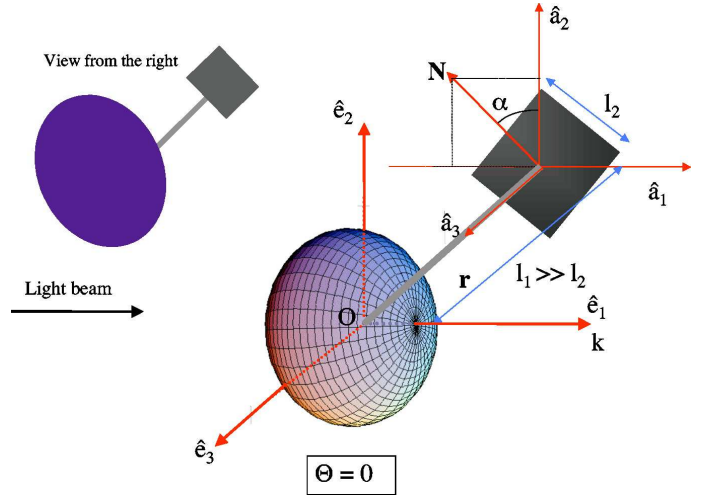}
\caption{Sketch of an AMO consisting of a weightless mirror attached to a perfectly reflecting ellipsoid. Principal axes $\hat{a}_{1},\hat{a}_{2}$ and $\hat{a}_{3}$ define a body coordinate system, and $\hat{e}_{1},\hat{e}_{2}$ and $\hat{e}_{3}$ define the lab coordinate system. $\Theta$ is the angle between $\hat{a}_{1}$ and the radiation direction $\bk \|{\hat{e}_{1}}$. ${\bf N}$ is the normal vector lying in the plane $\hat{a}_{2}\hat{a}_{3}$, and $\alpha$ is the tilted angle of the mirror in the body coordinate system. ${\bf r}$ is the radius vector of length $l_{1}$ connecting the ellipsoid to the mirror, $l_{2}$ is the mirror size and $l_{1}\gg l_{2}$ is assumed.} 
\label{f1}
\end{figure}

The geometry of AMO is depicted in Figure \ref{f1}. It has a weightless mirror tilted by an angle $\alpha$ about the axis ${\bf a}_{2}$ attached to an ellipsoid. Detail comparisons of RATs on the model and irregular grains suggested $\alpha=\pi/4$. For a given direction of the grain with respect to the radiation direction $\bk$, RATs arising from the scattering of photons on the mirror are functions of the angles $\Theta, \beta$ and $\Phi$. LH07a calculated three components of RATs at $\Phi=0$: $Q_{e_{1}}(\Theta, \beta, 0), Q_{e_{2}}(\Theta, \beta, 0)$ and $Q_{e_{3}}(\Theta, \beta, 0)$. 

LH07a found that whether the RAT alignment has high-J or low-J attractors depends on the ratio of torque components $q^{max}=Q_{e_{1}}^{max}/Q_{e_{2}}^{max}$ (see Appendix A). Therefore, this parameter has been used to characterize the RAT alignment in LH07a. For irregular grains, LH07a found that $q^{max}$ depends on the wavelength of radiation, grain shape, size and composition, ranging from $10^{-3}$ to $10^{2}$. Thus, we treat $q^{max}$ as a variable parameter in this paper.

To conform with numerical calculations using DDSCAT, we combine the functional forms of RATs from AMO with the scaling for the magnitude of RATs as a function of wavelength and grain size. The final RATs used in our series of papers are given in Appendix A. We note that for the AMO, the functional forms of RATs do not depend on wavelength, so that we can simplify the notations by using ${\bf Q}_{\Gamma}$ instead of $\overline{\bf Q}_{\Gamma}$.

\subsection{Torque-free motion}
The motion of a triaxial grain in the absence of external torques, namely, torque-free motion, is well-known (see Landau \& Lifshitz 1976; also WD03). Let consider an irregular grain with angular momentum $J$, and rotational energy $E$. It is convenient to introduce a dimensionless parameter\footnote{This parameter was denoted by $q$ in WD03, however, we use p to avoid the possible confusion with $q^{max}$.}
\bea
p=\frac{2I_{1}E}{J^{2}},\label{p}
\ena
where $I_{1}$ is the inertia moment about the axis of maximum moment of inertia $\ba_{1}$, and other inertia moments are smaller than $I_{1}$, i.e., $I_{1}>I_{2}>I_{3}$. Here $p$ spans in the range $1$ to $I_{1}/I_{3}$, where the lower and upper limits correspond to the grain rotation about the axis of maximum moment of inertia $\ba_{1}$ and axis of minimum moment of inertia $\ba_{3}$, respectively.

For the torque-free motion, the angular velocity components about the three principal axes, $\omega_{1}, \omega_{2}$ and $\omega_{3}$ can be obtained analytically as functions of $p$ and time $t$ (see WD03; HL08a for their expressions).
\begin{figure}
\includegraphics[width=0.5\textwidth]{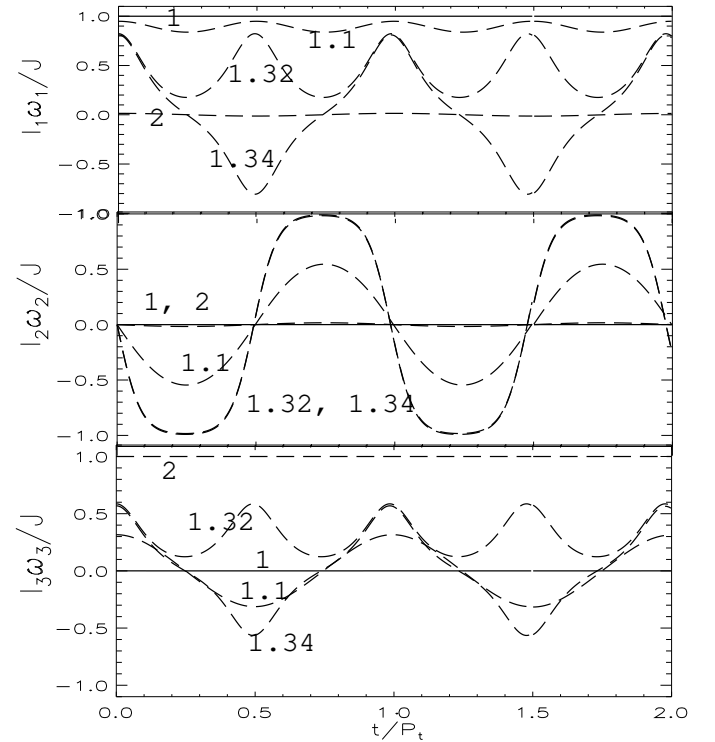}
\caption{Angular velocity components as functions of time $t/P_{t}$ with $P_{t}$ being the rotation period, for different values of $p$ for an irregular grain with $I_{1}:I_{2}:I_{3}=2:1.5:1$ and in a positive flipping state, i.e., $\omega_{1}$ and $\omega_{3}$ both being positive for $p<I_{1}/I_{2}$. The value of $p$ corresponding to each curve is labelled.} 
\label{omega}
\end{figure}
In Figure \ref{omega} we plot $\omega_{1}, \omega_{2}$ and $\omega_{3}$ as functions of time $t/P_{t}$ with $P_{t}$ being the rotation period, for different $p$ for an irregular grain with ratio of inertia moments $I_{1}:I_{2}:I_{3}=2:1.5:1$. The plot corresponds to a positive flipping state, i.e., $\omega_{1}$ and $\omega_{3}$ are both positive for $p<I_{1}/I_{2}$. Similarly, when $\omega_{1}$ and $\omega_{3}$ are both negative for $p<I_{1}/I_{2}$, we call negative flipping state (see HL08a for details).

We can see that when $p\rightarrow 1$ or $I_{1}/I_{3}=2$, the rotation is stable about $\ba_{1}$ and $\ba_{3}$ axes, respectively. However, the rotation is not stable about the intermediate axis $\ba_{2}$ when $p=1.32$ and $1.34$, which are very close to $I_{1}/I_{2}=1.333$. In fact, the middle panel shows that as $p\rightarrow I_{1}/I_{2}$, the grain gets into a negative state with $I_{2}\omega_{2}/J=-1$ for almost one half period, and then reverses its rotational state, staying there for another half period. Of course, at $p=I_{1}/I_{2}$, we have $I_{2}\omega_{2}/J=1$ and $\omega_{1}=\omega_{3}=0$.

Because the timescale of torque-free motion is much shorter than the internal relaxation timescale and other dynamical timescales (e.g., Larmor precession, gas damping times), we can average RATs over such a motion to obtain $\tilde{Q}_{\Gamma}(\Theta, \beta,\Phi)$ where the tilde denotes averaging over torque-free motion. Resulting RATs are functions of $\xi,\phi,\psi,p$ and $J$.

\subsection{Radiative torques in alignment coordinate system}

To study the alignment of the angular momentum ${\bf J}$ with the magnetic field ${\bf B}$, we represent RATs in the spherical coordinate system $J, \xi, \phi$, the so-called alignment coordinate system (see the lower panel in Fig. \ref{f2}).

In this alignment coordinate system, the RAT $\bG_{rad}$ can be written as
\bea
\bG_{rad}=M[F(\xi,\phi,p,J)\hat{\bf \xi}
  +G(\xi,\phi,p,J) \hat{\bf \phi}+H(\xi,\phi,p,J)\hat{\bf J}],\label{eq8}
\ena
where the dependence on the angle between the radiation direction ${\bf k}$ and ${\bf B}$, $\psi$, is omitted,
\bea
M=\frac{\gamma {u}_{rad}a^{2}\bar{\lambda}}{2},\label{M}
\ena
 $F$, which is the torque component parallel to $\hat{\bf \xi}$, acts to change the orientation of ${\bf J}$ with respect to ${\bf B}$. $H$, the component parallel to $\hat{\bf J}$, is to spin grains up and $G$ induces the precession of ${\bf J}$ about the magnetic field or radiation. The later are given by 
\bea
H(\xi,\phi,p,J)&=&\tilde{Q}_{e_{1}}(\xi, \phi,p,J)(\mbox{cos }\psi \mbox{cos }\xi -\mbox{sin }\psi \mbox{sin  }\xi \mbox{cos }\phi)\nonumber\\
&&+\tilde{Q}_{e_{2}}(\xi, \phi,p,J)(\mbox{sin}\psi \mbox{cos }\xi + \mbox{cos }\psi\mbox{sin }\xi\mbox{cos  }\phi)\nonumber\\
&& +\tilde{Q}_{e_{3}}(\xi, \phi,p,J)\mbox{sin }\xi \mbox{sin }\phi, \label{eq6c} \\
F(\xi,\phi,p,J)&=&\tilde{Q}_{e_{1}}(\xi, \phi,p,J)(-\mbox{sin }\psi \mbox{cos }\xi \mbox{cos }\phi-\mbox{sin  }\xi \mbox{cos }\psi)\nonumber\\
&&+\tilde{Q}_{e_{2}}(\xi, \phi,p,J)(\mbox{cos }\psi \mbox{cos }\xi \mbox{cos }\phi-\mbox{sin }\xi
\mbox{sin }\psi)\nonumber\\
&&+\tilde{Q}_{e_{3}}(\xi, \phi,p,J)\mbox{cos }\xi \mbox{sin }\phi,\label{eq6a}
\ena
where $\xi, \psi$ and $\phi$ are angles describing the direction of $\bJ$ in the lab coordinate system (see Fig. \ref{f2}, lower panel). Here $\tilde{Q}_{e_{i}}(\xi,\phi,p,J)$  with $i=1,2$ and $3$ are RAT components along $\be_{i}$ axes, respectively, obtained by averaging $Q_{e_{i}}$ over torque-free motion.

We define also the torque components along grain axes $\ba_{1}, \ba_{2}$ and $\ba_{3}$ as
\bea
Q_{a_{i}}(\xi,\phi,p,J)={\bf Q}_{\Gamma}.{\bf a}_{i} \mbox{~for~ i=1 to 3},
\ena
and thus, the work done by RATs per second
\bea
Q_{\omega}={\bf Q}_{\Gamma}.{\bf \omega}=Q_{a_{1}}\omega_{1}+Q_{a_{2}}\omega_{2}+Q_{a_{3}}\omega_{3}.\label{qome1}
\ena
Here $\omega_{i}$ for $i=1,2$ and $3$ corresponds to the components of angular velocity along three grain axes $\ba_{i}$. Their expressions for torque-free motion are given in Appendix A of HL08a. In addition, from Figure \ref{f2}, we have
\bea
\ba_{1}&=&{\bf e}_{1}\mc\Theta+{\bf e}_{2}\ms\Theta\mc\Phi+{\bf e}_{3}\ms\Theta\ms\Phi,\\
\ba_{2}&=&-{\bf e}_{1}\ms\Theta\mc\beta+{\bf e}_{2}(\mc\Theta\mc\Phi\mc\beta-\ms\Phi\ms\beta)\nonumber\\
&&+{\bf e}_{3}(\mc\Theta\ms\Phi\mc\beta+\mc\Phi\ms\beta),\\
\ba_{3}&=&{\bf e}_{1}\ms\Theta\ms\beta-{\bf e}_{2}(\mc\Theta\mc\Phi\ms\beta+\ms\Phi\mc\beta)\nonumber\\
&&+{\bf e}_{3}(-\mc\Theta\ms\Phi\ms\beta+\mc\Phi\mc\beta).
\ena

\begin{figure}
\includegraphics[width=0.5\textwidth]{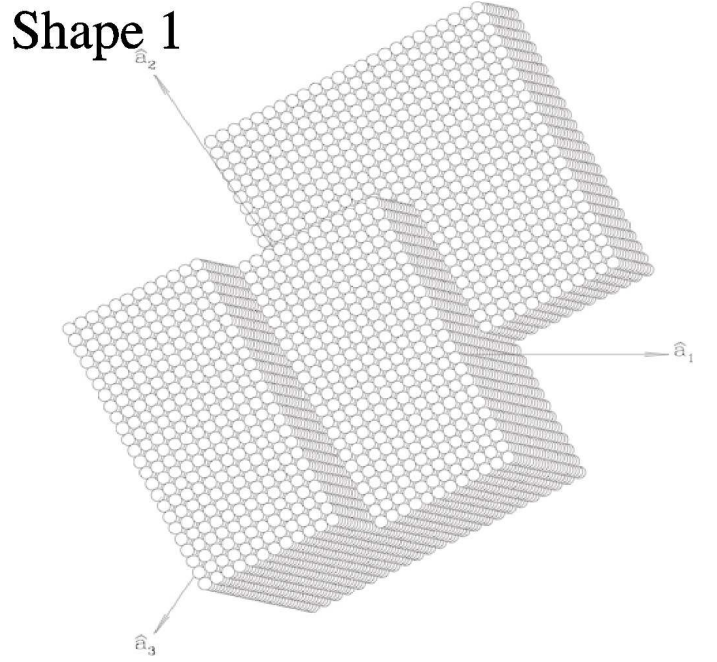}
\caption{Sketch of the irregular grain, shape 1, built from 13 cubes adopted in our calculations. The inertia moments along principal axes are $I_{1}:I_{2}:I_{3}=1.745:1.61:0.876$.}
\label{f3}
\end{figure}

\subsection{Calculations of RATs}

Taking analytical expressions $Q_{e_{i}}(\Theta,\beta,\Phi=0)$ with $i=1,2$ and $3$ from AMO (see Appendix A), we first calculate $Q_{e_{i}}(\Theta,\beta,\Phi)$  by using equations (\ref{aeq4})-(\ref{aeq6}) for $\Theta\in [0,\pi]$, $\beta\in[0,2\pi]$, and $\Phi\in[0,2\pi]$.
 Next, we calculate RATs for $J$ from $0.1$ to $20 I_{1}\omega_{T}$, $\xi$ from $0$ to $\pi$, $\phi$ from $0$ to $2\pi$, and $p$ from $1$ to $I_{1}/I_{3}$ and for a given light direction $\psi$. For a given set of parameters $J, \xi, \phi, p$, we use coordinate transformations to get angles $\Theta,\beta$ and $\Phi$ as functions of Euler angles (see Appendix A in HL08a). Finally, torque components $Q_{e_{i}}(\xi,\psi,\phi)$ are obtained using interpolation for 
$Q_{e_{i}}(\Theta,\beta,\Phi)$. Then we average $Q_{e_{i}}(\xi,\psi,\phi)$ over Euler angles, i.e., averaging over torque-free motion, to get ${\tilde Q}_{e_{i}}(\xi,\psi,\phi)$. 

Substituting ${\tilde Q}_{e_{i}}(\xi,\psi,\phi)\rangle$ into equations (\ref{eq6c}), (\ref{eq6a}) and (\ref{qome1}) we obtain $F(\xi,\psi,\phi), H(\xi,\psi,\phi)$ and $Q_{\omega}(\xi,\psi,\phi)$.  We then average them over angle $\phi$ to get the Larmor precession averaged values. Since these torques are functions of $J,\xi,p$, we denote them by $\langle F\rangle_{\phi}(J,\xi,p), \langle H\rangle_{\phi}(J,\xi,p)$ and $\langle Q_{\omega}\rangle_{\phi}(J,\xi,p)$, where the dependence on $\psi$ is omitted. We will use these results to solve equations of motion for the alignment in \S 4.

\section{Time scales: gas damping, internal relaxation and radiative alignment}
\subsection{Gas damping time}
We represent the magnitude of torques in terms of the thermal angular momentum, and the gas damping time. The later, for the sake of simplicity, can be obtained for an oblate spheroid, with the moments of inertia $I_{\|}=I_{1}, I_{\perp}=I_{2}=I_{3}$.\footnote{For the oblate spheroid with major and minor axes $a$ and $b$, $I_{\|}=2Ma^{2}/5=(8\pi\rho/15)ba^{4}$, and $I_{\perp}=(4\pi\rho/15) a^{2}b(a^{2}+b^{2})$, but sometimes we define the equivalent sphere with the same volume of the grain, and use $I_{i}=(8\pi/15)\rho \alpha_{i} a^{5}$ with $\alpha_{i}$ for $i=1,2$ and $3$ being dimensionless factors (see WD03).} The expressions for triaxial ellipsoids are mor complex, but do not change results substantially. Then, the thermal angular momentum is given by
\bea
J_{th}=\sqrt{I_{\|}k_{B}T_{gas}}=\sqrt{\frac{8\pi\rho a^{5}s}{15}k_{B} T_{gas}},\nonumber\\
=5.89\times 10^{-20} a_{-5}^{5/2}\hat{s}^{1/2}\hat{\rho}^{1/2} \hat{T}_{gas}^{1/2},\label{jtherm}
\ena
with $\rho$ density of material within the grain and $\hat{\rho}=\rho/3{~\mbox g cm}^{-3}$, $a$ grain size, $a_{-5}=a/10^{-5}{~\mbox cm}$, $\hat{s}=s/0.5$ with $s=b/a$ and $T_{gas}$ gas temperature and $\hat{T}_{gas}=T_{gas}/100$K. And the damping time due to gas collision (see Roberge, DeGraff \& Flatherty 1993, hereafter RDF93) is
\bea
t_{gas}=\frac{3}{4\sqrt{\pi}}\frac{I_{\|}}{n_{H}m_{H} a^{4}v_{th}\Gamma_{||}},\nonumber\\
=2.3\times 10^{12} a_{-5}\hat{s}\hat{T}_{gas}^{-1/2}\left(\frac{30 \mbox{ cm}^{-3}}{n_{H}}\right) {~\mbox s},\label{tgas}
\ena
where $v_{th}$ is the thermal velocity of a gas particle with density $n_{H}$, $m_{H}$ is the hydrogen mass, $\Gamma_{\|}$ is the geometrical parameter which is unity for sphere (see RDF93), and we adopted standard parameters in the ISM (see Table 1 in HL08a). 

\subsection{Internal relaxation timescales}
A rotating paramagnetic grain experiences internal dissipation of energy due to Barnett and nuclear relaxation (see Purcell 1979 and Lazarian \& Drain 1999b, hereafter LD99b, respectively). In addition, superparamagnetic inclusions can increase significantly the rate of internal relaxation (Lazarian \& Hoang 2008). 

Disregarding other relaxation processes within the grain, e.g., inelastic relaxation (Purcell 1979; Lazarian \& Efroimsky 1998) as well as the relaxation due to superparamagnetic inclusions, the internal relaxation rate arising from Barnett and nuclear relaxation for a brick of $2a\times 2a\times 2b$ sides (see Fig. \ref{f15}, lower panel) is then
\bea
t_{int}^{-1}&\approx&t_{Bar}^{-1}+t_{nucl}^{-1},\nonumber\\
&=&1.310^{-7}\hat{\rho}^{-2}\hat{s}[0.5+0.125\hat{s}^{2}]^{2}a_{-5}^{-7}\hat{s}\hat{T}_{gas}\hat{T}_{d}^{-1}\nonumber\\
&&\times\left(\frac{J}{I_{1}\omega_{T}}\right)^{2}(h-1)f(J).\label{tint}
\ena
Here the Barnett and nuclear relaxation times  $t_{Bar}$ and $t_{nucl}$ are taken from Hoang \& Lazarian (2008b; HL08b), $s=b/a$, $h=I_{\|}/I_{\perp}$, $\omega_{T}=\left(2kT_{gas}/I_{1}\right)^{1/2}$ is the thermal angular velocity of grain at the gas temperature $T_{gas}$, $\hat{T}_{d}=T_{d}/15$K with $T_{d}$ being the dust temperature, and
\bea
f(J)=1.610^{5}[1+\left(\frac{\omega_{1}\tau_{n}}{2}\right)^{2}]^{-2}+0.47[1+\left(\frac{\omega_{1}\tau_{el}}{2}\right)^{2}]^{-2}.
\ena
For grains larger than $1\mu$m, $f(J)\approx 1.6\times 10^5$ for $J=J_{th}$.
In above equations, $\omega_{1}=J\mc\theta/I_{\|}$, and $\tau_{n}$ is given by
\bea
\tau_{n}&=&1/(\tau_{ne}^{-1}+\tau_{nn}^{-1})
\ena  
where $\tau_{ne}=3\times 10^{-4}\left(2.7/g_{n}\right)^{2}\left(10^{22} cm^{-3}/n_{e}\right)$ s and $\tau_{nn}=\hbar/(3.8g_{n}n_{n}\mu_{n})\approx 0.58 \tau_{ne}\left(n_{e}/n_{n}\right)$ with electron and nucleus density $n_{e}$ and $n_{n}$ are the relaxation time of interaction nucleus-electron and nucleus-nucleus spins, respectively. Also, $\tau_{el}=2.9\times10^{-11}$s is the spin-spin relaxation time of electronic spins, and $\mc\theta=1/2$ is chosen (see LD99b; HL08b)

\subsection{Radiative alignment timescale}
The characteristic timescale for RATs to accelerate the grain from $J=0$ to $J=I_{1}\omega$ is estimated by
\bea
t_{rad}=\frac{I_{1}\omega}{\Gamma_{rad}}=\frac{I_{1}\omega}{M\langle H\rangle_{\phi}},
\ena
where $\langle H\rangle_{\phi}$ is the spin-up component of RATs, that is averaged over the torque-free motion and the precession angle $\phi$ using equation (\ref{eq6c}).

Using $M$ given by equation (\ref{M}) and $I_{1}=16\rho a^{5}s/3$ for the brick of $2a\times 2a\times 2b$ sides, we obtain
\bea
t_{rad}=\frac{2^{3/2}(16\rho k T_{gas}/3)^{1/2}a^{1/2}s^{1/2}}{\bar{\lambda} u_{rad}\gamma\langle H\rangle_{\phi}}.
\ena
For standard parameters of the ISM, it yields
\bea
t_{rad}\approx2.8\times 10^{10}\hat{\rho}\hat{T}_{gas}\hat{s}a_{-5}^{1/2}\frac{1.2~\mu m}{\bar{\lambda}}\frac{u_{ISRF}}{u_{rad}}\frac{10^{-3}}{\gamma\langle H\rangle_{\phi}} \mbox{ s},\label{trad}
\ena
where $\hat{s}=s/0.5$ with $s=b/a$ being the ratio of long to short axes, $u_{ISRF}$ is the energy density of interstellar radiation field (hereafter ISRF; see Mathis, Merger, \& Panagia 1983).

\subsection{Radiative alignment time versus internal relaxation time}
To compare the effect of RATs and internal relaxation, let us estimate the ratio of their timescales. Following equations (\ref{tint}) and (\ref{trad}) we obtain for a brick grain
\bea
\frac{t_{rad}}{t_{int}}&=& 3.7\times 10^{3}\hat{\rho}^{-1.5}[0.5+0.125\hat{s}^{2}]^{2}{a}_{-5}^{-6.5}\frac{\hat{T}_{gas}^{1.5}}{\hat{T}_{d}}\frac{1}{\hat{u}_{rad}\hat{\lambda}}\nonumber\\ 
&&\times \frac{10^{-3}}{\gamma\langle H\rangle_{\phi}}
\left(\frac{\omega}{\omega_{T}}\right)^{3}f(J),\label{eq12a1}
\ena
Here $\hat{T}_{d}=T_{d}/15 \mbox{K}$, $\hat{u}_{rad}={u}_{rad}/u_{ISRF}$, and $\hat{\lambda}=\bar{\lambda}/1.2~\mu m$. $u_{ISRF}=8.64\times 10^{-13} \mbox{erg cm}^{-3}$ is the energy density of interstellar radiation field (hereafter ISRF; see Mathis, Mezger \& Panagia 1983), and $\gamma=0.1$.

Using the scaling of RAT magnitude obtained in LH07a: $\langle H\rangle_{\phi} \approx  |Q_{\Gamma}| \approx 0.4\left(\frac{\bar{\lambda}}{a}\right)^{-3}$ for $\overline{\lambda}> 1.8 a$. We get
\bea
\frac{t_{rad}}{t_{int}}&=& 8.4\times 10^{5}\hat{\rho}^{-1.5}[0.5+0.125\hat{s}^{2}]^{2}\hat{a}_{-5}^{-9.5}\frac{\hat{T}_{gas}^{1.5}}{\hat{T}_{d}}\frac{\hat{\lambda}^{2}}{\hat{u}_{rad}}\nonumber\\
&&\times\left(\frac{\omega}{\omega_{T}}\right)^{3}f(J).\label{eq12a}
\ena
For $\overline{\lambda}<1.8 a$, $\langle H\rangle_{\phi} \sim 0.4$, 
\bea
\frac{t_{rad}}{t_{int}}= 70\hat{\rho}^{-1.5}\hat{s}\hat{a}_{-5}^{-6.5}\frac{\hat{T}_{gas}^{1.5}}{\hat{T}_{d}}\frac{1}{\hat{u}_{rad}\hat{\lambda}}\left(\frac{\omega}{\omega_{T}}\right)^{3}f(J).\label{eq12b}
\ena
 Both the increase of the grain size and the radiation intensity lead to the decrease of $t_{rad}$.

\begin{figure}
\includegraphics[width=0.49\textwidth]{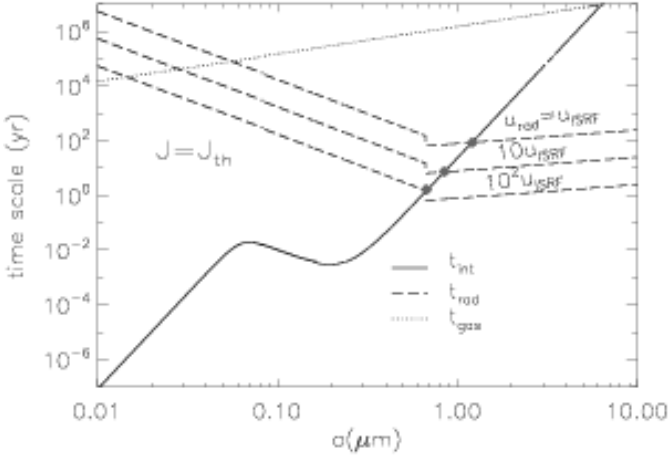}
\caption{Timescales as functions of grain size for internal relaxation ($t_{int}$), radiative ($t_{rad}$) and gas damping ($t_{gas}$) calculated for $J=J_{th}$ and the ISM. Filled circles denote where $t_{int}=t_{rad}$.}
\label{f2b}
\end{figure}

For a grain size $a=1.2~\mu$m, $f(J=J_{th})\sim 10^{5}$, so it can been seen from equation (\ref{eq12b}) that  $t_{rad}/t_{int}\sim 0.67$ for the ISRF (see also Fig. \ref{f2b}). 

Figure \ref{f2b} shows $t_{gas}$, $t_{int}$ and $t_{rad}$ as functions of $a$ for the ISM obtained from equations (\ref{tgas}), (\ref{tint}) and (\ref{trad}), respectively. It can be seen that for grains smaller than $\sim 1\mu m$, $t_{int}/t_{rad}\ll 1$, and decreases steeply with $a$ decreasing. For grains larger than $\sim 1.2\mu m$, we see that $t_{int}/t_{rad}>1$, and $t_{int}/t_{rad}$ increases rapidly with $a$. When the mean energy density $u_{rad}$ increases by$10^{2}$ time, the size corresponding to $t_{int}=t_{rad}$ decreases from $1.5$ to $0.6\mu$m.

Note that earlier works on grain alignment dealt with the alignment of interstellar grains with size in the range from $0.005$ to $0.25\mu$m. For this range of grain size, internal relaxation is very strong, so that the average of RATs over thermal fluctuations arising from internal relaxation was accounted for (WD03; HL08a). In some circumstances, e.g. accretion disks, molecular clouds, where larger grains corresponding to weak internal relaxation are expected, we need to study the internal and external alignment at the same time.

Another characteristic timescale involving in the grain dynamics is the Larmor precession time of grain magnetic moment about an ambient  magnetic field.
Due to the Barnett effect, a rotating paramagnetic grain develops a magnetic moment $\mu_{Bar}$ which is proportional to the angular velocity (see Dolginov \& Mytraphanov 1976). The value of $\mu_{Bar}$ is given by 
\bea
\mu_{Bar}=\frac{\chi(0)V\hbar}{g\mu_{B}}\omega,\label{mu}
\ena
where $\chi(0)=4.2\times 10^{-2}f_{p}\hat{T}_{d}^{-1}$ with $f_{p}$ being the fraction of paramagnetic material, is the magnetic susceptibility at zero frequency, $V$ is the volume, $g$ is gyromagnetic ration, which is $\sim 2$ for electrons, and $\mu_{B}=e/2m_{e}$ is the Bohr magneton.

The Larmor precession time in an external magnetic field $B$ is then
\bea
t_{L}=\frac{2\pi I_{1}\omega}{\mu_{Bar} B}\approx4.2\times 10^{5}a_{-5}^{2} \frac{\hat{\rho}\hat{T}_{d}}{\hat{B}\hat{\chi}(0)}\mbox{ s},
\ena
where $\hat{B}=B/(5 \mu G)$ and $\hat{\chi}(0)=\chi(0)/10^{-3}$, and we used $I_{1}=8\pi\rho a^{5}/15$ for inertia moment.

It is easy to see that this time scale is shorter than both internal relaxation and radiative alignment timescales for grains larger than about $\mu$m (see Fig.~\ref{f2b}), so that we can average RATs over the Larmor precession while dealing with the overall alignment.

\section{RAT alignment in the absence of internal relaxation}

Now let us study grain dynamics for grains larger than $1 \mu m$ in the diffuse interstellar medium. For this range of grain size, $t_{int}>t_{rad}$, the influence of internal relaxation to grain alignment can be disregarded. As a result, we follow both the evolution of angular momentum and grain axes subject to RATs.

\subsection{Equations of motion}

The orientation of ${\bf J}$ with respect to the magnetic field ${\bf B}$ is described by the following equation
\bea
\frac{d{\bf J}}{dt}={\bf \Gamma}_{rad}-\frac{{\bf J}}{t_{gas}},\label{eq16a}
\ena
where ${\bf \Gamma}_{rad}$ is the vector of radiative torque (see eq. \ref{a1}), $t_{gas}$ is the gas damping time.

 With the use of equation (\ref{eq8}), we can rewrite equation (\ref{eq16a}) in the spherical coordinate system $J, \xi, \phi$, assuming that both precession timescales of ${\bf a}_{1}$ around ${\bf J}$ and ${\bf J}$ around ${\bf B}$ are smaller than the internal relaxation time:
\bea
\frac{dJ}{dt}&=&M(\langle H(J,\xi,p)\rangle_{\phi})-\frac{J}{t_{gas}},\label{eq18}\\
\frac{d\xi}{dt}&=&\frac{M}{J}\langle F(J,\xi,p)\rangle_{\phi},\label{eq19}
\ena
where $\langle H(J,\xi,p)\rangle_{\phi}$ and $\langle F(J,\xi,p)\rangle_{\phi}$ are spin-up and aligning components, respectively, obtained from averaging equation (\ref{eq6c}) and (\ref{eq6a}) over torque-free motion, and the precession angle $\phi$ and $M$ is given by equation (\ref{M}). Also, we assume that the gas damping is isotropic, and we disregard the alignment by paramagnetic dissipation.

Unlike the axisymmetric grains where the angle $\theta$ between $\ba_{1}$ and $\bJ$ is constant during torque-free motion, irregular grains exhibit the torque-free wobbling, and the conserved quantities are $E$, the total energy and $J$, the value of angular momentum. 
Therefore, for an irregular grain, similar to WD03, we can use $p$ to describe the dynamical evolution of grain axes. With the use of equation (\ref{p}), and taking time derivative of $p$ we obtain
\bea
\frac{J^{2} dp}{dt}&=&M[I_{1}\langle Q_{\omega}(J,\xi,p)\rangle_{\phi}-pJ\langle H(J,\xi,p)\rangle_{\phi}]\nonumber\\
&&-\frac{J(p-1)}{t_{int}}\left(\frac{1-pI_{3}/I_{1}}{1-I_{3}/I_{1}}\right),\label{eq20b}
\ena
where $\langle Q_{\omega}\rangle$ is the average of torque component along the angular velocity given by equation (\ref{qome1}), and $t_{int}$ is given by equation (\ref{tint}) (see Appendix B). Let us define, 
\bea
\langle K(J,\xi,p)\rangle_{\phi}=\langle Q_{\omega}(J,\xi,p)\rangle_{\phi}/(pJ),\label{kmean}
\ena
then, equation (\ref{eq20b}) becomes
\bea
\frac{J dp}{dt}&=&Mp[\langle K(J,\xi,p)\rangle_{\phi}-\langle H(J,\xi,p)\rangle_{\phi}]\nonumber\\
&&-\frac{(p-1)}{t_{int}}\left(\frac{1-pI_{3}/I_{1}}{1-I_{3}/I_{1}}\right),\label{eq20c}
\ena

Before studying grain dynamics induced by RATs, we first want to consider how the torque components $\langle F(J,\xi,p)\rangle_{\phi}$, $\langle H(J,\xi,p)\rangle_{\phi}$ and $\langle K(J,\xi,p)\rangle_{\phi}$ change with $p$ and $\xi$. We adopt AMO that has the inertial property of  triaxial ellipsoids with ratio of inertia moments $I_{1}:I_{2}:I_{3}=2:1.5:1$ and $I_{1}:I_{2}:I_{3}=1.745:1.610:0.8761$. The later is identical with the ratio of inertia moments of the shape 1 (see Fig. \ref{f3}) and naturally the parameter $q^{max}$ can be changed for the AMO (LH07a). A $1.5\mu$m grain size and interstellar radiation field are adopted in this section.

\subsection{Variation of RATs with $p$: stationary points for $p$}

\begin{figure}
\includegraphics[width=0.49\textwidth]{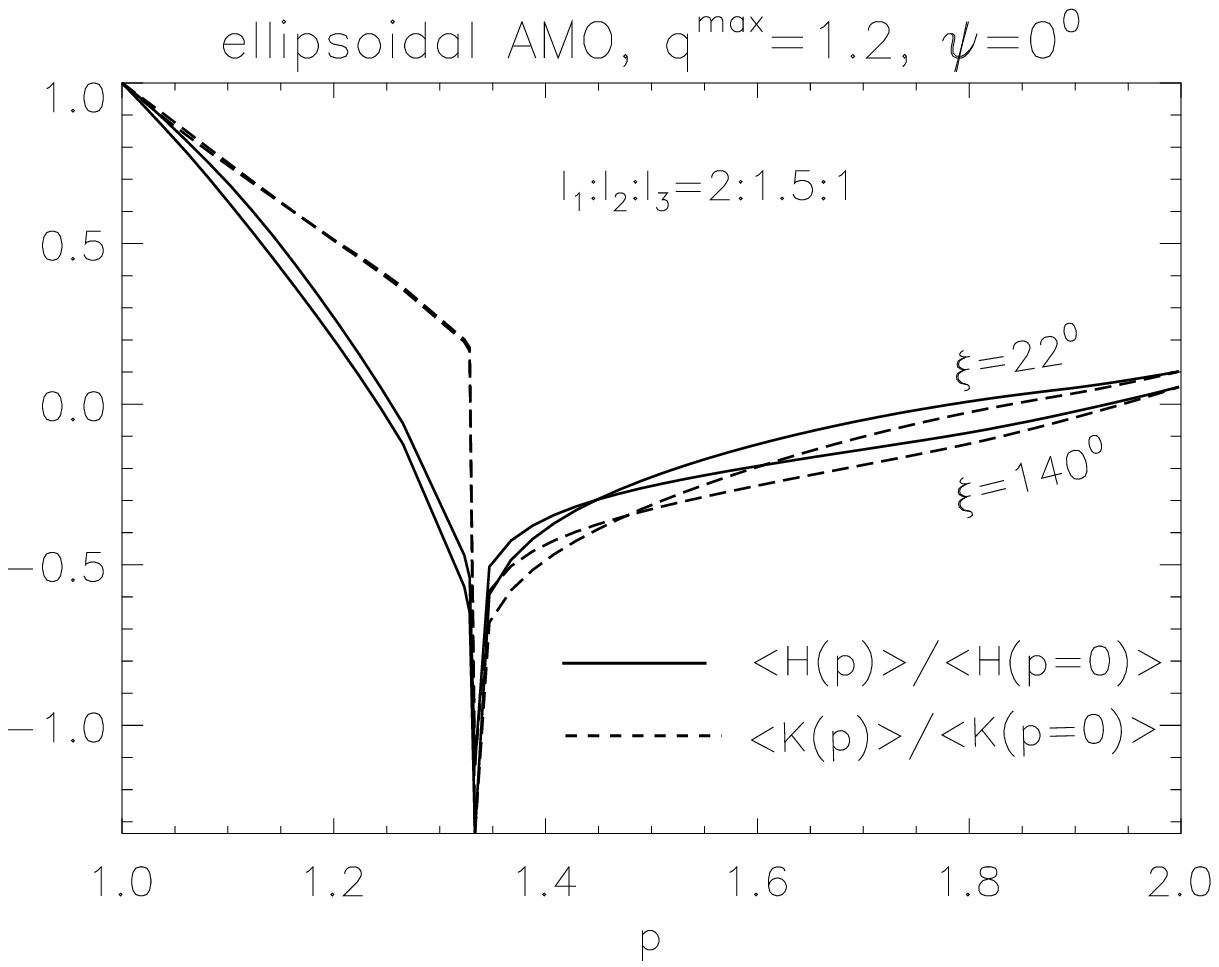}
\includegraphics[width=0.49\textwidth]{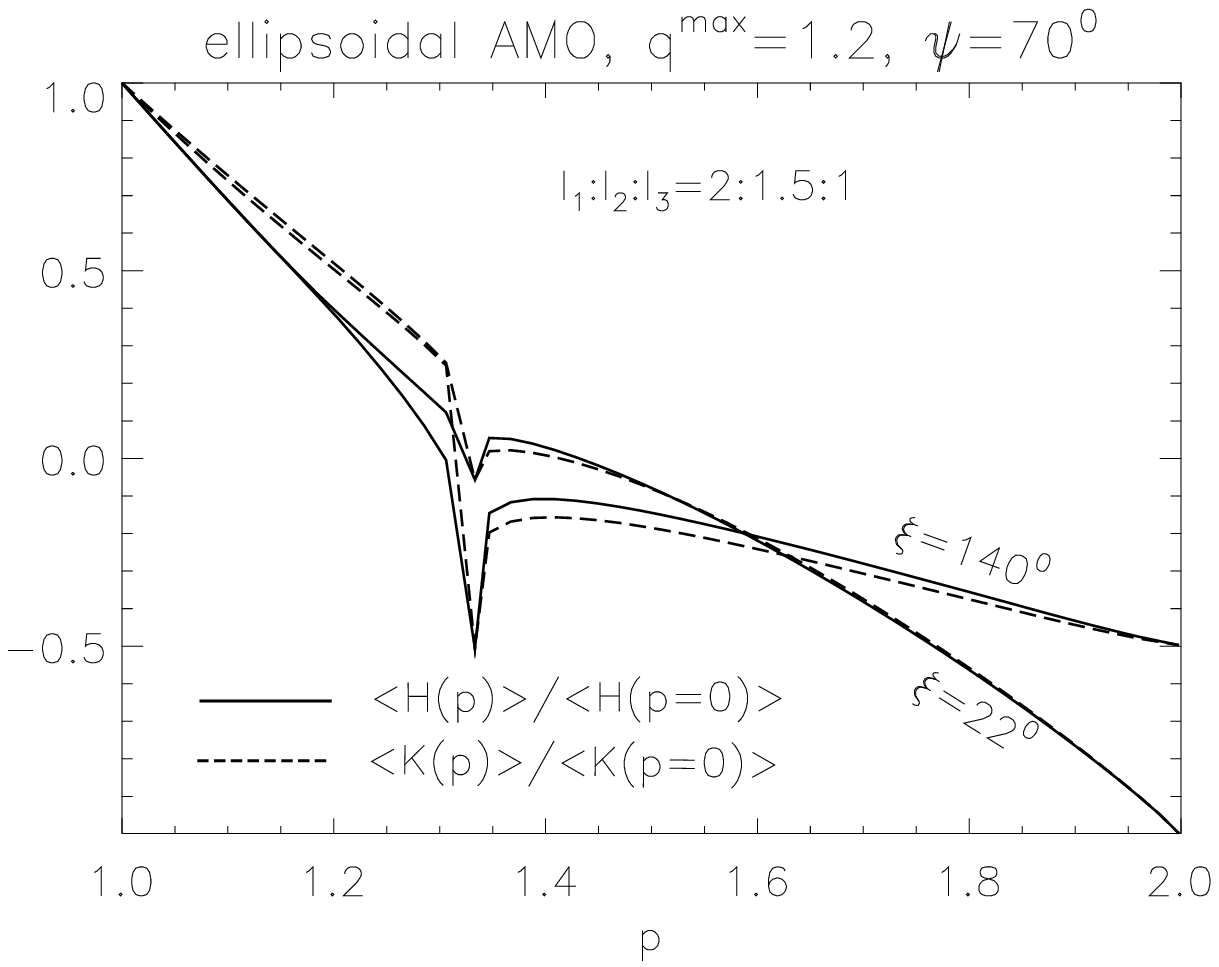}
\caption{Torque components averaged over torque-free motion,  $\langle H(p)\rangle$ and $\langle K(p)\rangle$, for different angles $\xi$ and for two radiation directions $\psi=0^{\circ}$ and $70^{\circ}$. These torques $\langle H(p)\rangle$ and $\langle K(p)\rangle$ coincide for $p=1,~I_{1}/I_{2}$ and $I_{1}/I_{3}$, corresponding to three stationary points 1, 2 and 3.}
\label{f6a}
\end{figure}

\begin{figure}
\includegraphics[width=0.49\textwidth]{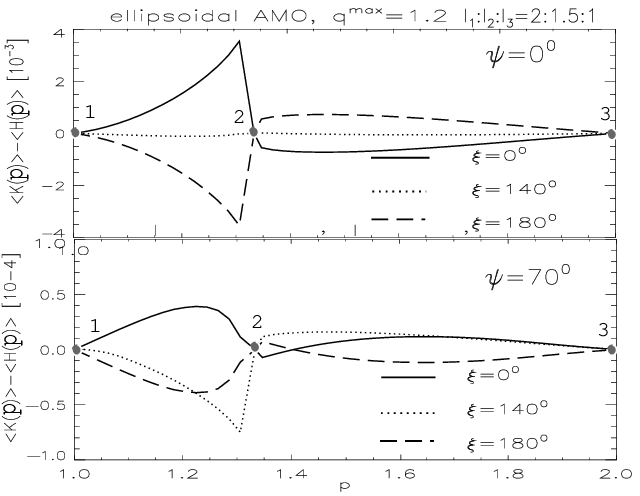}
\includegraphics[width=0.49\textwidth]{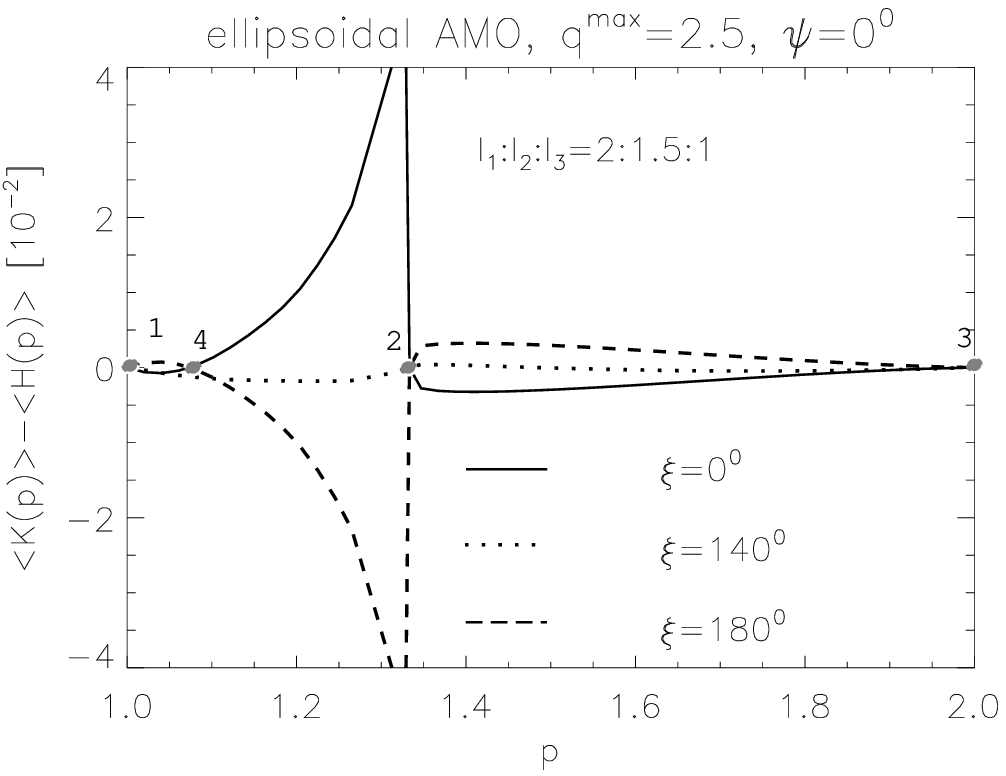}
\caption{Difference between $\langle K(p)\rangle$ and $\langle H(p)\rangle$ for different angles $\xi$ for $\psi=0^{\circ}$ and $70^{\circ}$ and $q^{max}=1.2$ (upper panel) and $ \psi=0^{\circ}, q^{max}=2.5$ (lower panel). Zero points denoted by filled circles corresponding to $\langle K\rangle-\langle H\rangle=0$ are stationary points of $p$. In the upper panel, three stationary points are 1,2 and 3, corresponding to $p=1,~I_{1}/I_{2}$ and $I_{1}/I_{3}$. In the lower panel, in addition to stationary points 1,2 and 3, there exists another stationary point 4.}
\label{f6b}
\end{figure}

In Figure \ref{f6a} we plot $\langle H(J,\xi,p)\rangle_{\phi}$ and $\langle K(J,\xi,p)\rangle_{\phi}$ as functions of $p$ for $J=10I_{1}\omega_{T}$, $\xi=22^{\circ},~140^{\circ}$ and $\psi=0$ (upper panel) and $70^{\circ}$ (lower panel) for the AMO with parameter $q^{max}=1.2$.

We see that $\langle H(J,\xi,p)\rangle_{\phi}$ and $\langle K(J,\xi,p)\rangle_{\phi}$ decrease as $p$ increases to $I_{1}/I_{2}$, and then increase as $p$ increases to $I_{1}/I_{3}$. This is because as $p \rightarrow I_{1}/I_{2}$, grain axes wobble more rigorously about $\bJ$, which results in the decrease of torques due to averaging over wobbling with large amplitude. As $p$ increases again, $\omega$ gets rotating stably about ${\ba}_{3}$, and averaged torques increase because the wobbling amplitude decreases.

In addition, $\langle H(J,\xi,p)\rangle_{\phi}$ and $\langle K(J,\xi,p)\rangle_{\phi}$ are identical at $p=1,~I_{1}/I_{2}=1.33$ and $I_{1}/I_{3}=2$. This arises from the fact that at these points, $\omega$ parallel to $\bJ$, so the projection of RATs onto $\bJ$, $\langle H(J,\xi,p)\rangle_{\phi}$, is the same with that on $\omega$, $\langle K(J,\xi,p)\rangle_{\phi}$.

The sharp decrease of $\langle K(J,\xi,p)\rangle_{\phi}$ when $p\rightarrow I_{1}/I_{2}$ is associated with the dynamics of a triaxial ellipsoid. Nevertheless, for $p$ very close to $I_{1}/I_{2}$, $\omega$ oscillates in the lab system, and gives rise to a sharp change of $\langle K(J,\xi,p)\rangle_{\phi}$ (see dashed lines in Fig.~\ref{f6a}).

Equation (\ref{eq20c}) shows that stationary points occur when $dp/dt=0$. In the absence of internal relaxation for large grains, it requires $\langle K(J,\xi,p)\rangle_{\phi}-\langle H(J,\xi,p)\rangle_{\phi}=0$.

Figure \ref{f6b} shows $\langle K(J,\xi,p)\rangle_{\phi}-\langle H(J,\xi,p)\rangle_{\phi}$ for the same AMO with $q^{max}=1.2$, $\psi=0^{\circ}, 70^{\circ}$ (upper) and $2.5$, $\psi=0^{\circ}$ (lower). There we omit the dependence of torques on $J$ and $\xi$. We remind the reader that these cases correspond to the alignment without and with high-$J$ attractors, respectively, when the perfect internal alignment of $\ba_{1}$ with $\bJ$ was assumed (see LH07a). In the upper panel, corresponding to the alignment without high-J attractors, there are three stationary points 1, 2 and 3. Let $p_{1},~p_{2}$ and $p_{3}$ be the value of $p$ at these points. It is easy to see that $p_{1}=1,~ p_{2}=I_{1}/I_{2}$ and $p_{3}=I_{1}/I_{3}$. In the alignment with high-J attractors (see lower panel), there exists another stationary point 4, and its value of $p$ is denoted by $p_{4}$. Our calculations show that $p_{4}$ depends on $q^{max}$ and $\psi$. In Figure \ref{f8a} we plot $q^{max}$ as functions of $p_{4}$ for two radiation directions $\psi=0$ and $70^{\circ}$. 

In addition, Figure \ref{f6b} shows that when $p\rightarrow I_{1}/I_{2}$, the figure exhibits very sharp changes. In fact, we see the fast changes of $\langle K(J,\xi,p)\rangle_{\phi}-\langle H(J,\xi,p)\rangle_{\phi}$ as $p$ increases from slightly smaller than $I_{1}/I_{2}$ to slightly larger than $I_{1}/I_{2}$. As $p=I_{1}/I_{2}$, it becomes zero.

\begin{figure}
\includegraphics[width=0.49\textwidth]{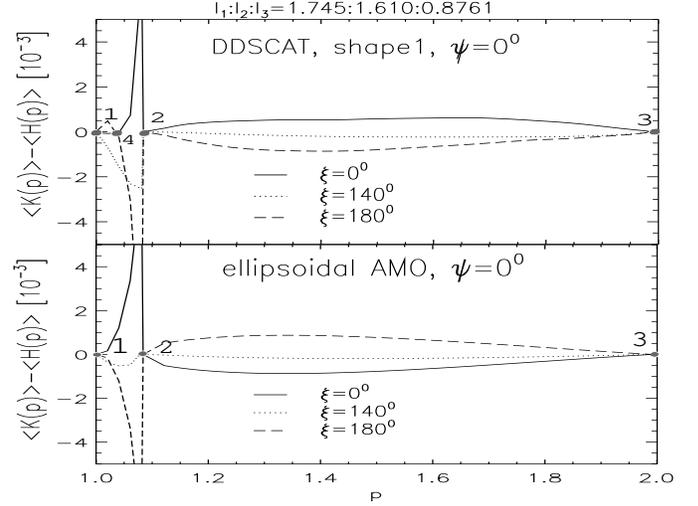}
\caption{Similar to Figure \ref{f6b} but for shape 1 and AMO with the same ratio of inertia moments $I_{1}:I_{2}:I_{3}=1.745:1.610:0.8761$ and $q^{max}=2.5$, and for $\psi=0^{\circ}$.}
\label{f6c}
\end{figure}

Similar to Figure \ref{f6b}, Figure \ref{f6c} shows $\langle K(J,\xi,p)\rangle_{\phi}-\langle H(J,\xi,p)\rangle_{\phi}$ for the shape 1 (see Fig.\ref{f3}) and AMO with the same ratio of inertia moments $I_{1}:I_{2}:I_{3}=1.745:1.610:0.8761$ and $q^{max}=2.5$, and for $\psi=0^{\circ}$. It can be seen that the AMO reproduces well the dependence of RATs on $p$ for the shape 1. For instance, we observe a good correspondence between the stationary points obtained, but the stationary point 4 in the AMO is very close to the point 1 in this case (point 4 is seen in the lower panel of Fig.7).

\subsection{Variation of RATs with $\xi$: stationary points for $\xi$}

\begin{figure}
\includegraphics[width=0.47\textwidth]{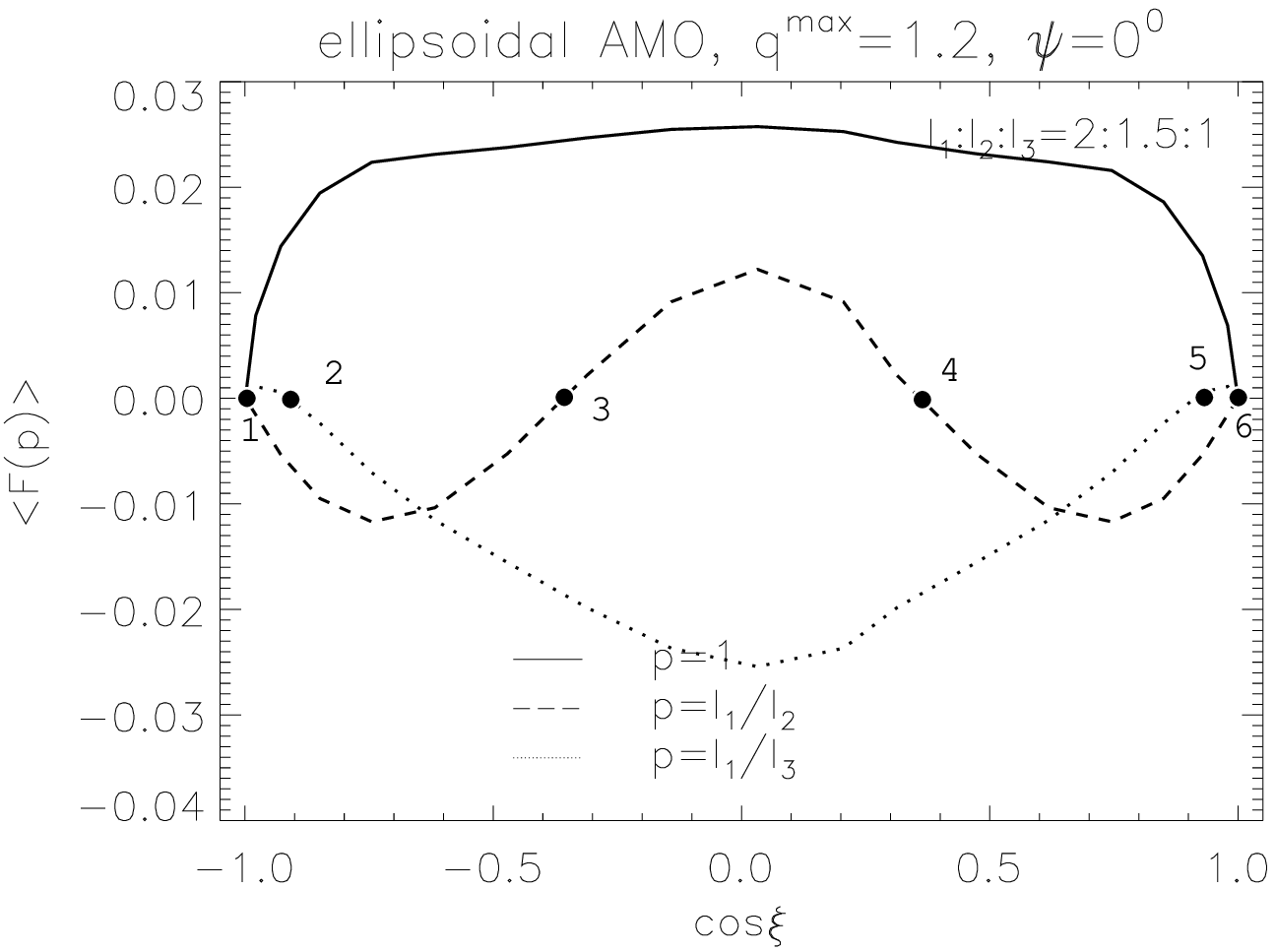}
\includegraphics[width=0.49\textwidth]{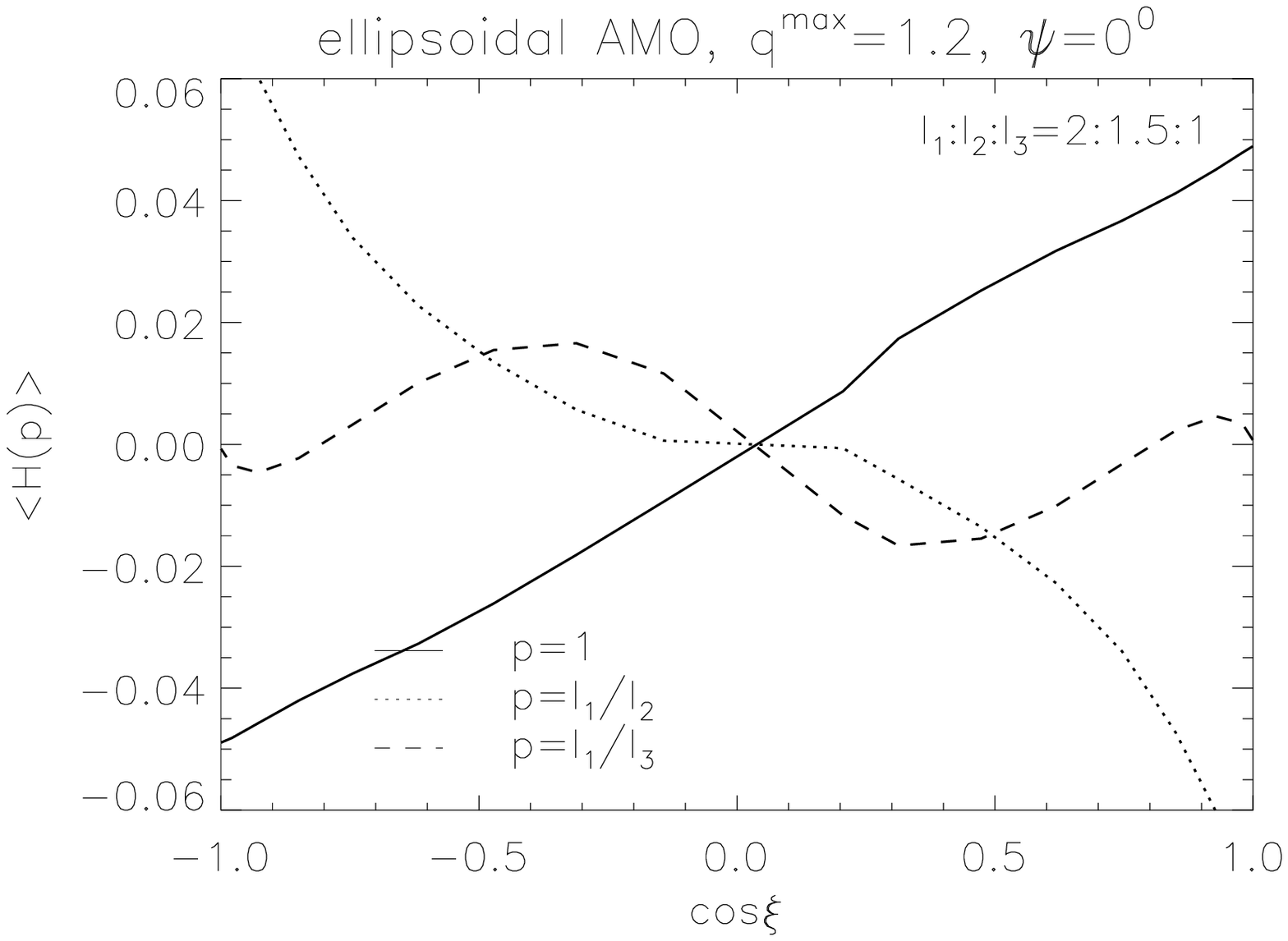}
\caption{Torque components $\langle F\rangle$ and $\langle H\rangle$ for $p=1, ~I_{1}/I_{2}$ and $I_{1}/I_{3}$. (1,6), (1,2,5,6) and (1,3,4,6) are stationary points for $p=1,~I_{1}/I_{2}$ and $I_{1}/I_{3}$, respectively.}
\label{f5a}
\end{figure}
\begin{figure}
\includegraphics[width=0.47\textwidth]{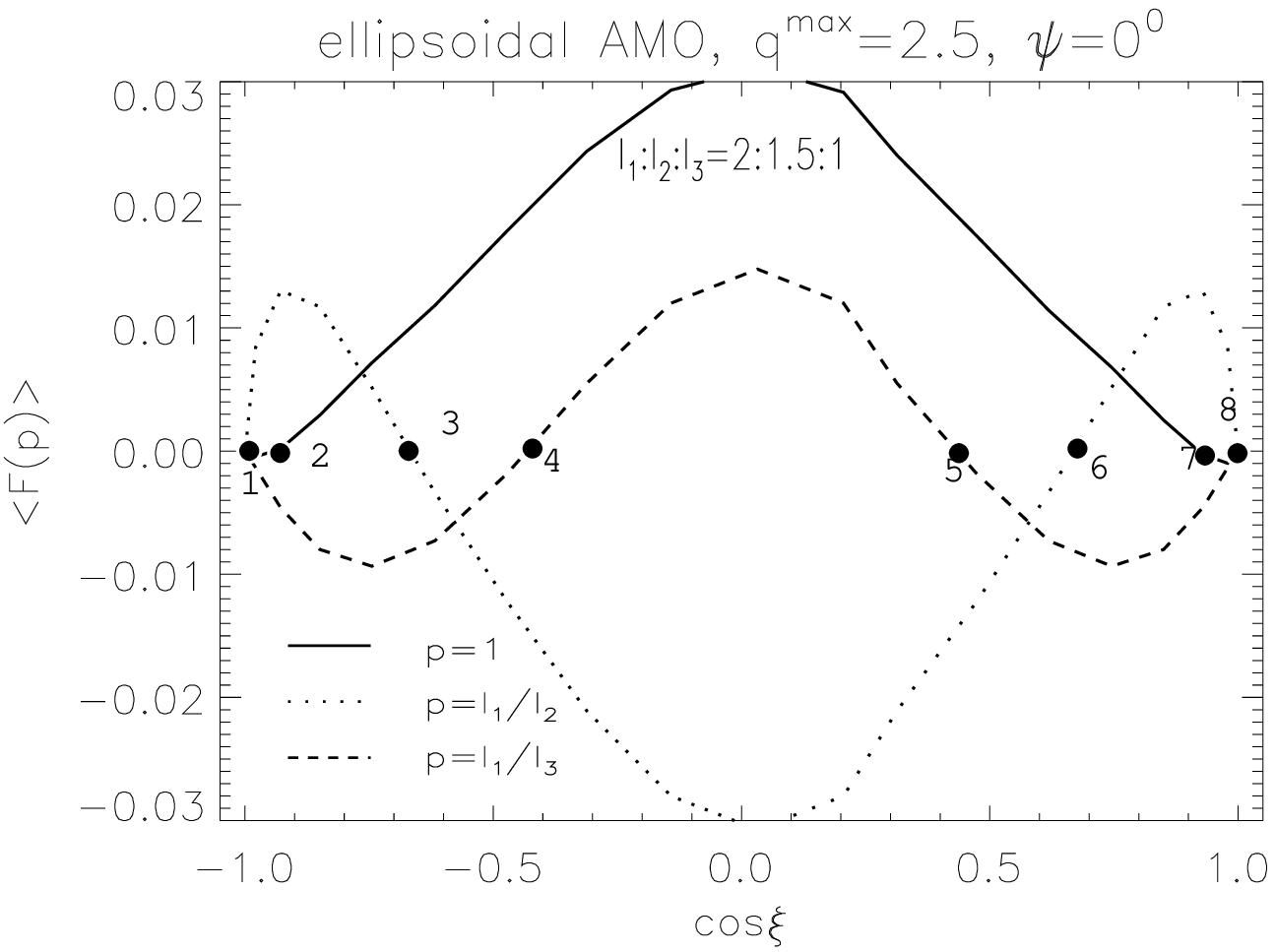}
\includegraphics[width=0.49\textwidth]{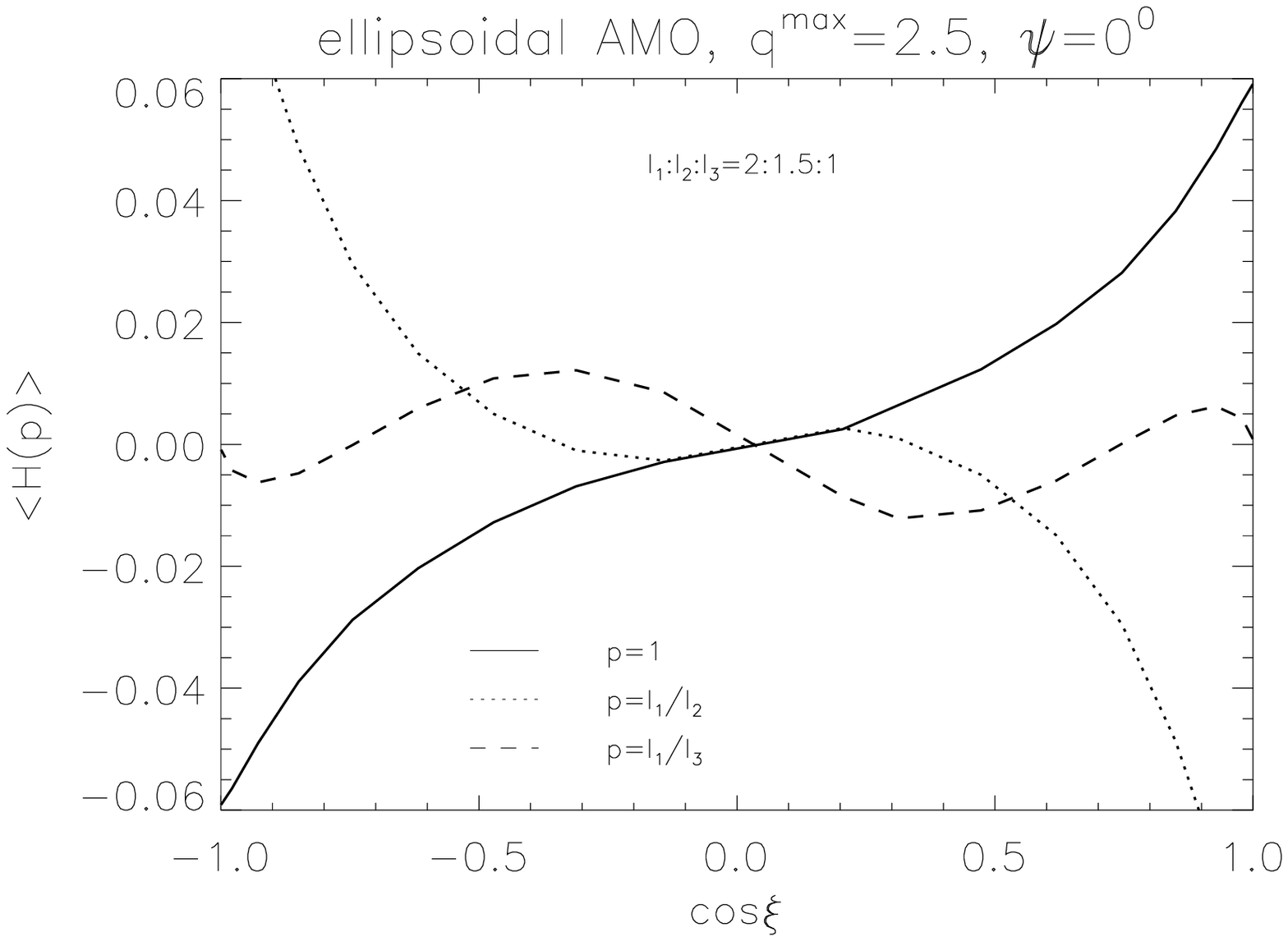}
\caption{Similar to Fig. \ref{f5a} but for the case $q^{max}=2.5$, i.e. alignment with high-J attractor for $\psi=0^{\circ}$. (1,2,7,8), (1,3,6,8) and (1,8) are stationary points for $p=1,~I_{1}/I_{2}$ and $I_{1}/I_{3}$, respectively.}
\label{f5b}
\end{figure}
Let us consider the forms of RATs as a function of $\xi$ at particular values $p=1$ (perfect coupling of $\ba_{1}$ with $\bJ$) and $I_{1}/I_{3}$ ($\ba_{1}\perp \bJ$) for the same AMO .

In Figure \ref{f5a} we present the torque components $\langle F(p)\rangle$ and $\langle H(p)\rangle$ for the alignment without high-J attractor, i.e., $q^{max}=1.2,~\psi=0^{\circ}$ for $p=1,~I_{1}/I_{2}$ and $I_{1}/I_{3}$. We see that stationary points $\xi=0$ and $\pi$, i.e., points 1 and 6, appear for every $p$. In addition to these points, for $p=I_{1}/I_{2}$ and $I_{1}/I_{3}$, there exist two other stationary points at $\mc\xi=\pm 9$ (2,5), and $\mc\xi=\pm 0.35$ (3,4), respectively. So, in the case of wrong internal alignment, $\bJ$ may not perfectly aligned with the magnetic field.

The upper panel of Figure \ref{f5b} is similar to that of Figure \ref{f5a}, but for the case of alignment with high-J attractor with $q^{max}=2.5$. For $p=1$, we see four stationary points (1,2,7,8), where (8) is a high-J attractor. Other stationary points (1,3,6,8) and (1,4,5,8) are for $p=I_{1}/I_{2}$ and $I_{1}/I_{3}$, respectively. 

\subsection{Helicity versus axis of rotation}

For $p=I_{1}/I_{3}$, the rotation of the grain is about the axis of minor inertia $\ba_{3}$. In this case, RATs depend on the angle between the grain rotation axis $\ba_{3}$ and the radiation direction $\bk$. This angle is also the angle between $\bJ$ and $\bB$, $\xi$, when $\psi=0^{\circ}$. Therefore, the dashed line in Figure \ref{f5a} represents RATs as a function of the angle between the grain rotation axis and $\bk$.

For $p=1$, grain rotates about $\ba_{1}$ described by the angle $\Theta$. Thus, the solid line describes RATs as a function of $\Theta=\xi$. We can see a large difference of $\langle F(p)\rangle$ when the rotation axis changes from the axis of maximum moment of inertia to axis of minor inertia. The lower panel is similar to the upper one, but it shows $\langle H(p)\rangle$. We see that the helicity of the grain for $p=1$ is similar to that for $p=I_{1}/I_{3}$. In other words, the helicity of the grain is identical for the rotation around the maximal and minimal axes.

\subsection{Trajectory map}
\begin{figure}
\includegraphics[width=0.49\textwidth]{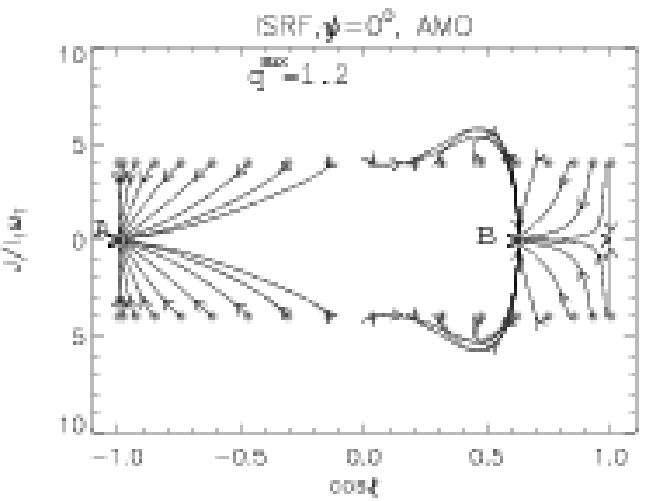}
\includegraphics[width=0.49\textwidth]{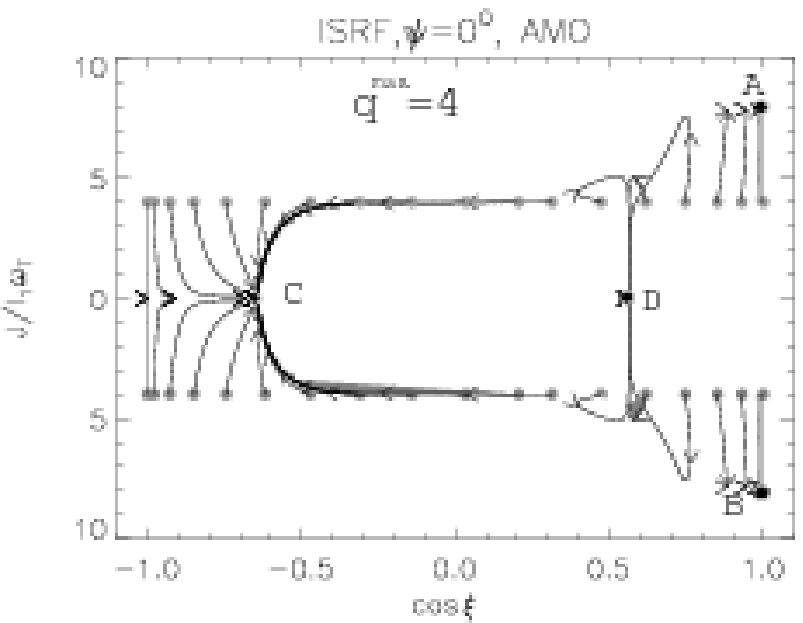}
\caption{Trajectory maps for the RAT alignment of a triaxial ellipsoid using the AMO, in the absence of internal relaxation for the ISRF with direction $\psi=0^{\circ}$ and with $q^{max}=1.2$ ({\it upper panel}) and $q^{max}=4$ ({\it lower panel}). A and B are low-$J$ attractors in the case $q^{max}=1.2$, but there are two high-$J$ attractors A and B for the case $q^{max}=4$ (lower panel; here the actual value of $J$ for A and B is $10^{2}$ times the value shown). }
\label{f6}
\end{figure}

To visualize grain alignment, we present trajectory maps using $J, \xi$ and $p$, which are the solutions of equations of motion (\ref{eq18})-(\ref{eq20c}). We adopt initial condition $J_{0}=4 I_{1}\omega_{T}$, $\xi_{0}$ spanning from $0$ to $\pi$, and $p_{0}=1.1$. Here we consider only one initial value $p_{0}<I_{1}/I_{2}$ (i.e., grain initially in positive flipping state).

In Figure \ref{f6} we present the trajectory maps for the RAT alignment of the triaxial AMO induced by the ISRF for $\psi=0$ and for $q^{max}=1.2$ and $4$, corresponding to the alignment without and with high-$J$ attractors (upper and lower panels). In the former case, the alignment occurs at two low-$J$ attractors A and B, where A corresponds to the perfect alignment of $\bJ$ with $\bB$, and B corresponds to $\mc\xi=0.64$. We see that the point B is the new low-$J$ attractor appearing in the absence of internal relaxation.\footnote{LH07a found that for the AMO with $q^{max}=1.2$, all grains are driven to one low-$J$ attractor A for $\psi=0^{\circ}$}.

In the case $q^{max}=4$, the alignment has a high-$J$ attractor A at $\mc\xi=1$, and low-$J$ attractors C and D with $\mc\xi=-0.65$ and $0.65$, respectively. 

In addition, we studied the radiative alignment for the case $q^{max}=0.78$ and $\psi=70^{\circ}$, which gives the trajectory map with high-$J$ attractor. We observed that the alignment is similar to the lower panel in Figure \ref{f6}.

\subsection{Internal alignment of grain axes with $\bJ$}
Let us discuss the internal alignment corresponding to the trajectory maps shown in Figure \ref{f6}. The initial value of $p$ was assumed to be smaller than  $I_{1}/I_{2}$.

For the alignment without high-$J$ attractor (the upper panel), the point A corresponds to the perfect internal alignment with $p=1$, while the point B has internal alignment with $\bJ$ along $\ba_{2}$ axis, which corresponds to $p=I_{1}/I_{2}$. 

For the case of alignment with high-$J$ attractor, i.e., $q^{max}=4$ and $\psi=0^{\circ}$ (lower panel of Fig. \ref{f6}), the internal alignment depends on the initial value of $p_{0}$. If $p_{0}<p_{4}$, then the lower panel of Figure \ref{f6b} shows that $\langle K(p)\rangle-\langle H(p)\rangle<0$ for $\mc\xi_{0}\sim 1$. As a result, after some time, $p$ decreases to $p=1$, i.e., perfect internal alignment, and the aligning torque $\langle F(p)\rangle$ approaches $\langle F(p=1)\rangle$. Therefore, the alignment returns to the case of perfect coupling of $\ba_{1}$ with $\bJ$ as studied in DW97 and LH07a, i.e., the alignment of $\bJ$ with $\bB$ is determined by $\langle F(p=1)\rangle$ and $\langle H(p=1)\rangle$. LH07a found that for the AMO with $q^{max}>2$ and $\psi=0^{\circ}$, the alignment has high-J attractors. We can see it in the lower panel of Figure \ref{f6} for $q^{max}=4$. Combined with the internal alignment, the high-J attractor corresponds to $\bJ\|\bB$ and long grain axes perpendicular to $\bB$. For $\mc\xi_{0} \sim -1$, we have $\langle K(p)\rangle-\langle H(p)\rangle>0$, so $p$ increases from $p_{0}$ to $p_{4}$ if $p_{0}<p_{4}$. Thus, the low-J attractor corresponds to the alignment with $p=p_{4}$. As a result, A and B correspond to the internal alignment with $\ba_{1}\|\bJ$, and C corresponds to internal alignment with $\ba_{1}\perp \bJ$. If $p_{0}>p_{2}$, then there are two attractors with $p=p_{4}$ and $p_{2}$.

\begin{figure}
\includegraphics[width=0.49\textwidth]{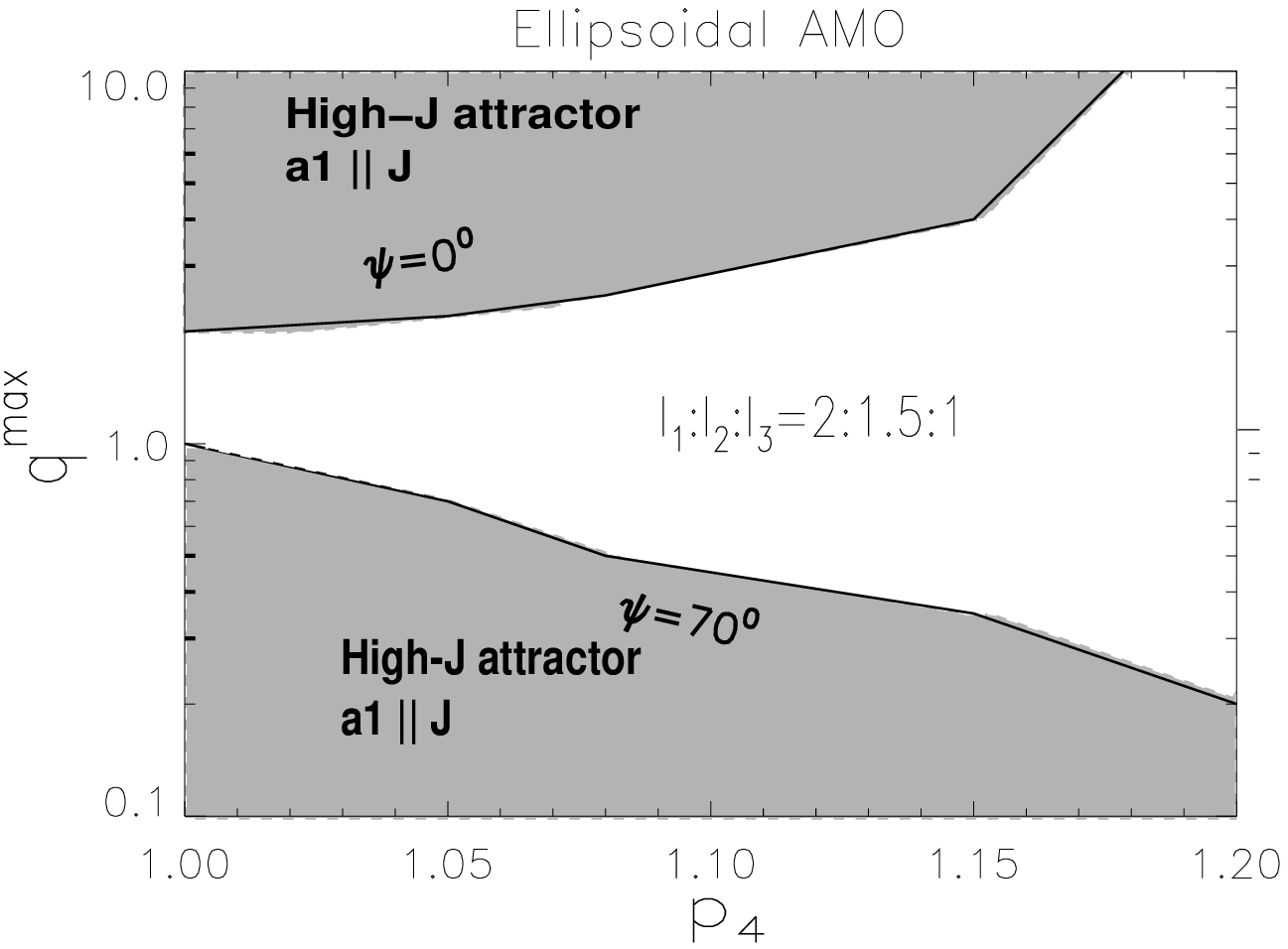}
\includegraphics[width=0.49\textwidth]{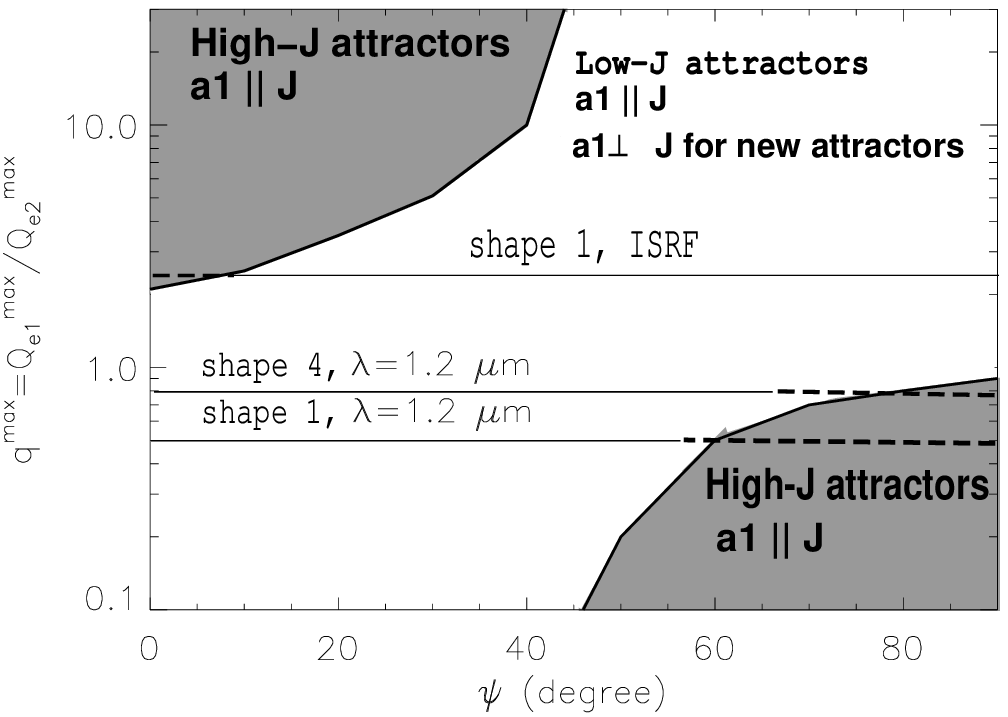}
\caption{{\it Upper:} Diagram showing the dependence of $p_{4}$-the value of $p$ at the stationary point 4 (see Fig. \ref{f6b} and \ref{f6c}) as function of $q^{max}$ for two light directions $\psi=0$ and $70^{\circ}$. Shaded areas represent the internal alignment when  the initial value, $p_{0}$ smaller than $p_{4}$. {\it Lower panel:} Diagram showing the alignment type depending on $q^{max}$ and $\psi$ when $p_{0}<p_{4}$.}
\label{f8a}
\end{figure}

Since the stationary point $p_{4}$ is important in determining the type of alignment, in the upper panel of Figure \ref{f8a} we plot $q^{max}$ against  $p_{4}$, for two light beam directions $\psi=0^{\circ}$ and $70^{\circ}$. Shaded areas describe the alignment with high-J attractor corresponding to a perfect alignment of $\bJ$ with ${\bf B}$ and perfect internal alignment, i.e., $\ba_{1}\|\bJ$.

The lower panel of Figure \ref{f8a} presents the alignment depending on $q^{max}$ and $\psi$. 

Moreover, when the internal alignment happens with $p=1$, then, we can predict the alignment for different $\psi$ if we know $q^{max}$. The result is presented in the lower panel of Figure \ref{f8a}. The solid curves, characterizing the boundary for the alignment with high-J attractor, are obtained assuming that $p_{0}<p_{4}$. We see that the alignment here is similar to what found in LH07a. We note that at $p=1$, the average over the wobbling for irregular grains is identical with the averaging over the rotation angle $\beta$ used in LH07a.

We note that the internal alignment with $p=I_{1}/I_{2}$ corresponds to the rotation of a grain along the axis of intermediate inertia moment $\ba_{2}$. For torque-free motion, this rotation is unstable. Therefore, gas bombardment can destabilize this internal alignment, and grains may return to the alignment with short or long axes perpendicular to $\bJ$.

\subsection{RAT alignment in presence of weak internal relaxation}
Let us consider a regime of weak internal relaxation, i.e., when $t_{rad}<t_{int}<t_{gas}$. Figure \ref{f2b} shows that grains with size in the range $1.5$ to $6\mu$m correspond to this situation. For this regime, it is convenient to introduce a new radiative timescale, $t'_{rad}$, which is defined by (using eq. \ref{eq20c})
\bea
t'_{rad}=\frac{I_{1}\omega}{M|\langle K(p)\rangle-\langle H(p)\rangle|_{max}},
\ena
where max denotes the maximal value of $|\langle K(p)\rangle-\langle H(p)\rangle|$ as a function of $p$ for given angles $\xi$ and $\psi$.

It can be seen that $t'_{rad}$ is about one order of magnitude larger than $t_{rad}$ given by equation (\ref{trad}). Therefore, it is possible to have $t'_{rad}>t_{int}>t_{rad}$. For this case, one interesting effect occurs with the wrong alignment previously discussed is that grains are driven to right alignment on a characteristic time $t_{int}$. However, for practical purposes, a change from $t_{rad}$ to $t'_{rad}$ is marginal due to a steep dependence.

\section{RAT alignment by dipole and quadrupole components of radiation field}

All earlier studies of the RAT alignment (DW96, DW97, WD03; LH07, LH08, HL08ab) have been done assuming that the radiation field can be described by a unidirectional beam directed along $\bk$ (see Fig. \ref{f2}) with a degree of anisotropy $\gamma$. This assumption is consistent with grains near a point radiation source. In many circumstances, e.g., in molecular clouds, and even in some diffuse clouds, the approximation of radiation field by dipole and quadrupole is more appropriate.

 Below, we study the RAT alignment induced by the dipole and quadrupole fields. For the sake of simplicity, we assume that the axis of maximum moment of inertia $\ba_{1}$ is constrained to be parallel to angular momentum $\bJ$ (i.e., thermal fluctuations and thermal flipping are completely disregarded; see DW97; LH07a). In fact, our study in HL08a proved that this is a sufficiently good approximation unless we want to calculate the exact degree of alignment.

\subsection{Coordinate systems}

\begin{figure}
\includegraphics[width=0.5\textwidth]{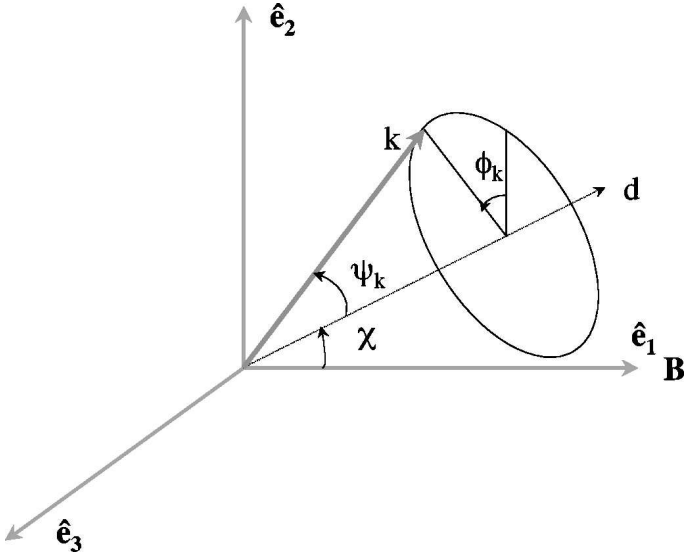}
\caption{The lab coordinate system $\be_{1}\be_{2}\be_{3}$ for the dipole field: $\bB$ is the magnetic field direction, ${\bf d}$ is the dipole axis, $\chi$ is the angle between the dipole axis ${\bf d}$ and the magnetic field $\bB$, ${\bf k}$ is the radiation direction. $\phi_{k}, \psi_{k}$ describe the direction of radiation ${\bf k}$ in the lab coordinate system.}
\label{f10}
\end{figure}

We define a lab coordinate system $\be_{1}\be_{2}\be_{3}$ ($\be_{i}$-system), with $\be_{1}$ parallel to ${\bf B}$, $\be_{2}\perp \be_{1}$, lies in a plane formed by the magnetic field $\bB$ and the dipole axis ${\bf d}$ for dipole field or symmetric axis for the quadrupole field, and $\be_{3}$ perpendicular to the plane $\be_{1}$ and $\be_{2}$ (see Fig. \ref{f10}).\footnote{The symmetric axis of the quadrupole field is defined by two dipole axes.} Let $\chi$ be the angle between ${\bf d}$ and $\bB$, and define a coordinate system $\be_{1}^{\chi}\|{\bf d}\be_{2}^{\chi}\be_{3}^{\chi}$ with $\be_{3}^{\chi}=\be_{3}$, so-called ${\bf d}$-system. Now, the radiation direction $\bk$ in the ${\bf d}$-system is determined by two angles $\psi_{k}$ and $\phi_{k}$ (see Fig. \ref{f10}). Since RATs depend on the angle between grain axes $\ba_{i}$ and the radiation direction $\bk$, let us define a coordinate system $\be_{i}^{0}$ with $\be_{1}^{0}\|\bk$, $\be_{2}^{0}$ lying in the plane $\bk$ and ${\bf d}$, and $\be_{3}^{0}$ perpendicular to this plane. The orientation of a grain in the $\be_{i}^{0}$ system is described by three angles $\Theta, \beta$ and $\Phi$ (see the upper panel of Fig. \ref{f2} with $\be_{i}$ are replaced by $\be_{i}^{0}$). The transformation between these coordinate systems are described in Appendix D.

\subsection{Dipole Component}
For a given direction of the grain in the lab coordinate system, the net RAT resulting from a dipole radiation field is given by
\bea
{\bf \Gamma}_{rad}=\sum_{k=1}^{N} {\bf Q}(\psi_{k})_{\Gamma}\frac{\gamma \bar{\lambda} a^{2} u_{d}(\psi_{k})}{2} \frac{\Delta \Omega_{k}}{2\pi},\label{eq12}
\ena
where $\psi_{k}$ is the angle between the radiation direction $\bk$ and the dipole axis ${\bf d}$, $\gamma$ is the degree of anisotropy, $u_{d}(\psi_{k})$ is the energy density of dipole field, $\Delta\Omega_{k}=\pi/N$ is the element of solid angle in the direction $\psi_{k}$, and the summation (\ref{eq12}) is performed in the range $\psi_{k}$ from $-\pi/2$ to $\pi/2$ (i.e., only outward radiation is accounted for). The spatial distribution of dipole radiation intensity is given by the usual expression 
\bea
u_{d}(\psi_{k})=u_{ISRF}\mbox{sin}^{2}\psi_{k}\label{eq13}.
\ena
We assume in equation (\ref{eq13}) that the energy density of the dipole field has the same amplitude to that of the ISRF, and we deal only with its spatial distribution. 

To study the RAT alignment, we need to know components of torques in the lab coordinate system, i.e., $Q_{e_{1}}(\xi,\psi,\phi), Q_{e_{2}}(\xi,\psi,\phi)$ and $Q_{e_{3}}(\xi,\psi,\phi)$. It is straightforward, but tedious to obtain $Q_{e_{i}}(\xi,\psi,\phi)$ from $Q_{\Gamma}(\Theta, \beta, \Phi)$ given by equations (\ref{a4})-(\ref{a5}) through a series of coordinate transformation. Firstly, we perform the transformation from $\bk$-system to ${\bf d}$-system, then, from ${\bf d}$-system to $\be_{i}$-system. Finally, we average the resulting torques in the spherical system $J, \xi, \phi$ over the azimuthal angles of $\bk$ about dipole axis, $\phi_{k}$, and the  Larmor precession angle $\phi$ (see Appendix E for more detail).

\begin{figure}
\includegraphics[width=0.5\textwidth]{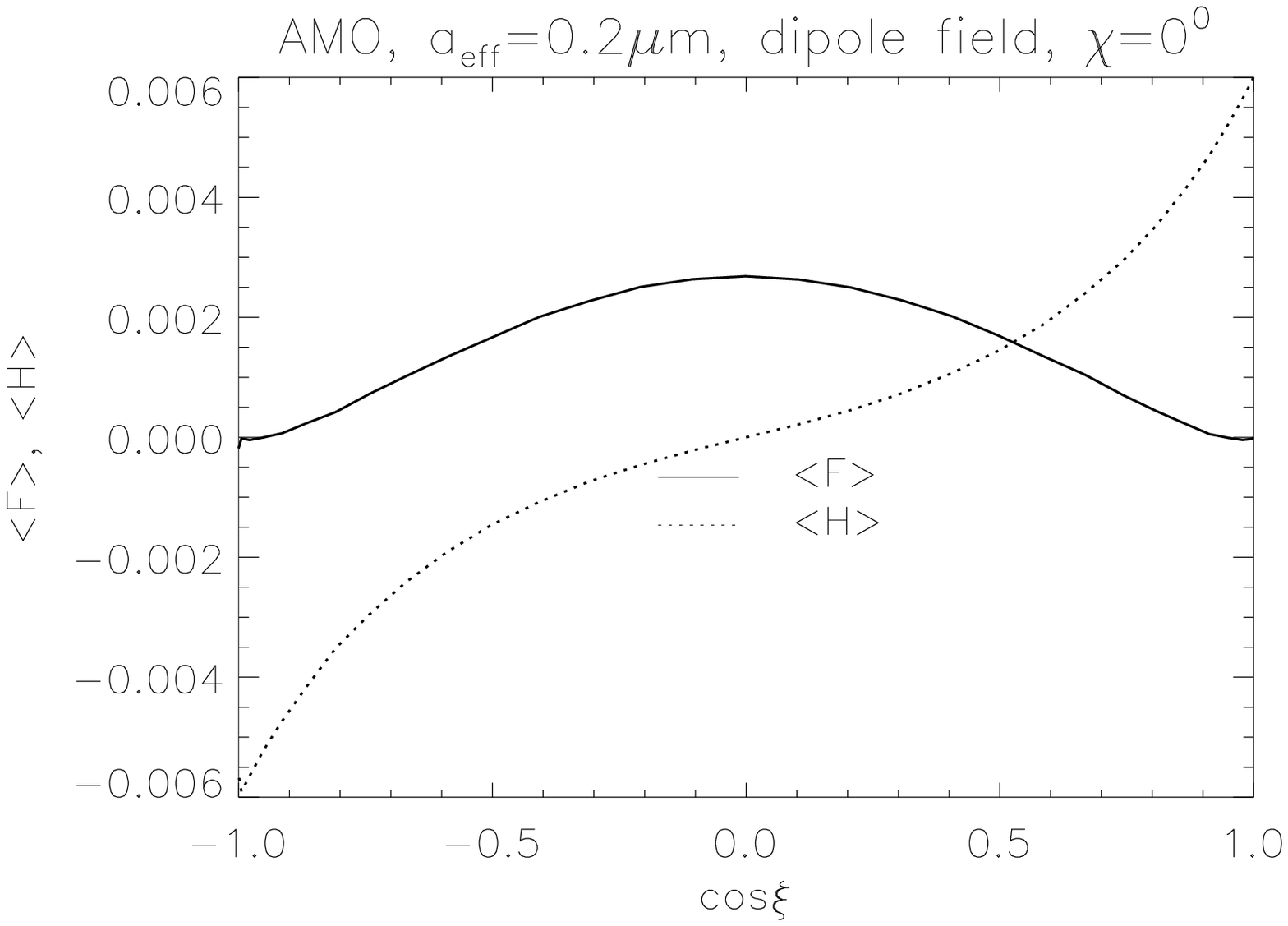}
\includegraphics[width=0.5\textwidth]{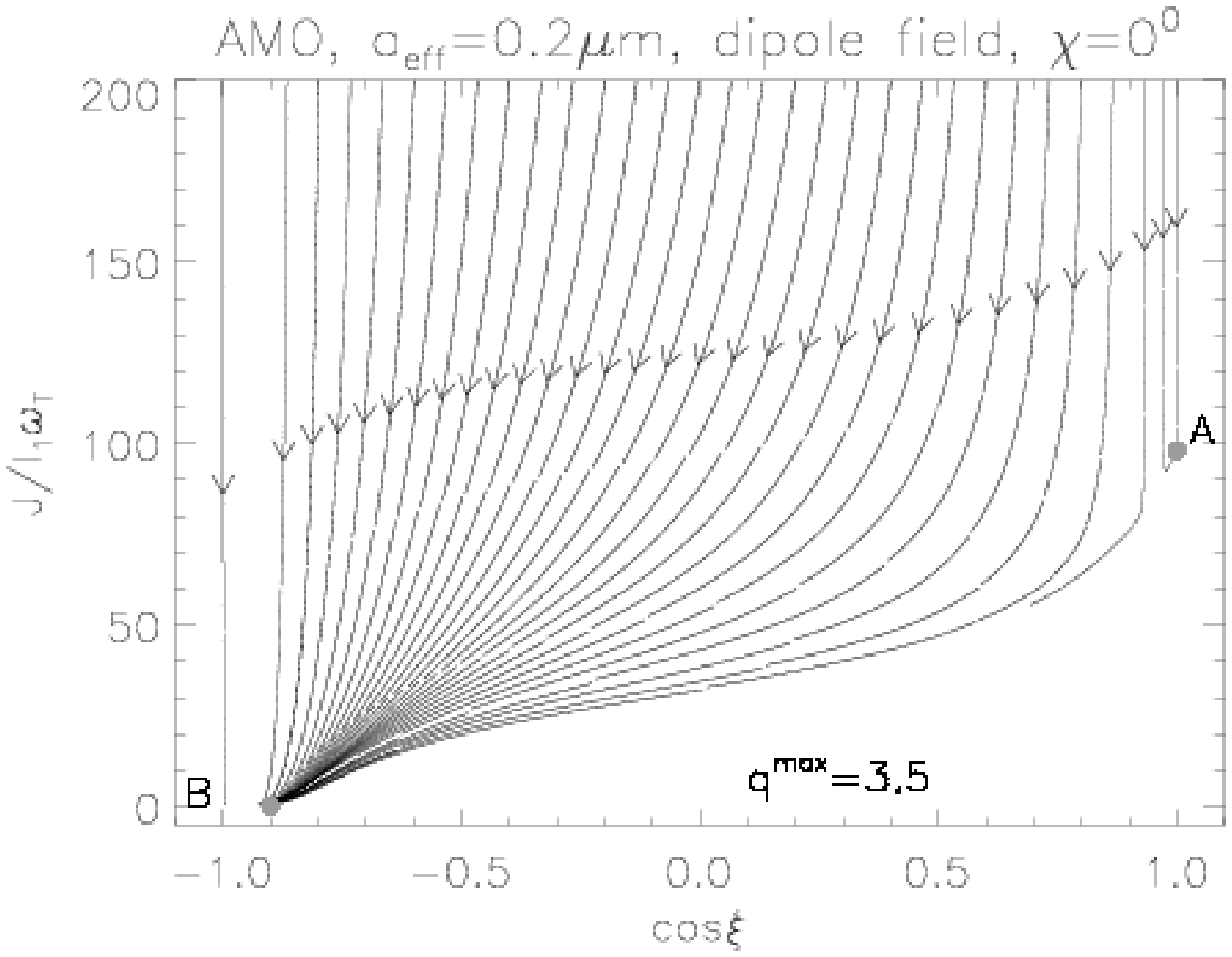}
\caption{{\it Upper panel:} Aligning, $\langle F\rangle_{\phi}$, and spin-up, $\langle H\rangle_{\phi}$ components for the AMO with $q^{max}=3.5$ induced by the dipole radiation field with $\chi=0$. {\it Lower panel:} The phase trajectory map obtained by torques in the upper panel. A is a high-$J$ attractor corresponding to the perfect alignment of $\bJ$ with respect to $\bB$, and B is a low-$J$ attractor.}
\label{f11}
\end{figure}

Using the obtained RATs, we solve the equations of motion to find $J, \xi$ as functions of time, and present their evolution in terms of phase trajectory maps (see LH07a for more detail). As an example, we use the AMO for a grain size $a=0.2 \mu m$, and $q^{max}=3.5$. For the dipole radiation field with $\chi=0^{\circ}$, the torque components and the corresponding trajectory map are shown in the upper and lower panel of Figure \ref{f11}, respectively. The upper panel shows the existence of stationary points at $\mc\xi\sim \pm 1$. It is possible to check that the stationary point $\mc\xi=1$ is an attractor because $\left.\langle F'\rangle_{\phi}/\langle H\rangle_{\phi}\right|_{\mc\xi=1}<0$ and $\langle H\rangle_{\phi}>0$. This high-$J$ attractor is denoted by a circle (A) in the trajectory map (see {\it lower panel}). In addition, there exists also a low-$J$ attractor B, as usual. Both stationary points correspond to a perfect alignment of angular momentum with the magnetic field.

\subsection{Quadrupole Component}
\begin{figure}
\includegraphics[width=0.5\textwidth]{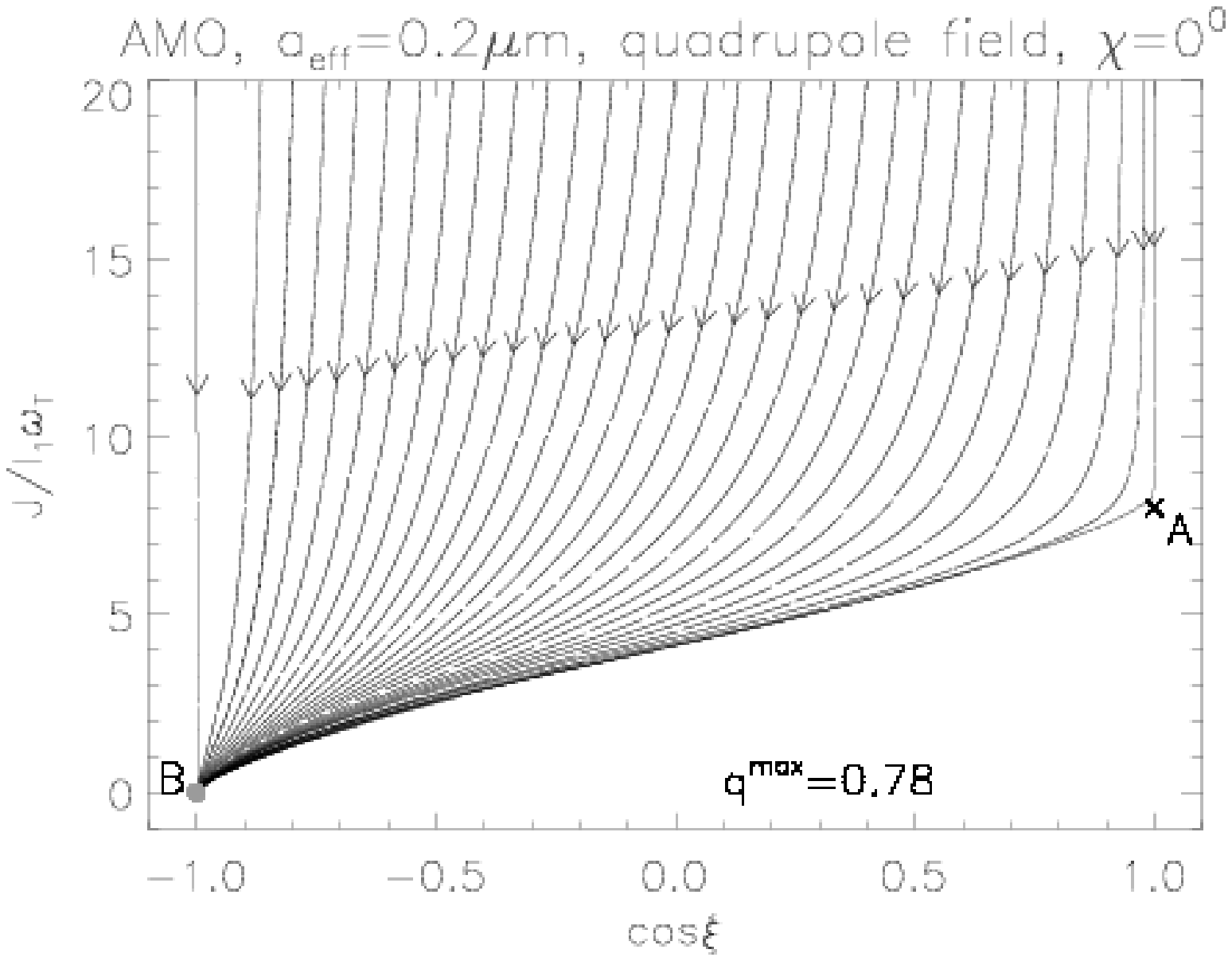}
\includegraphics[width=0.5\textwidth]{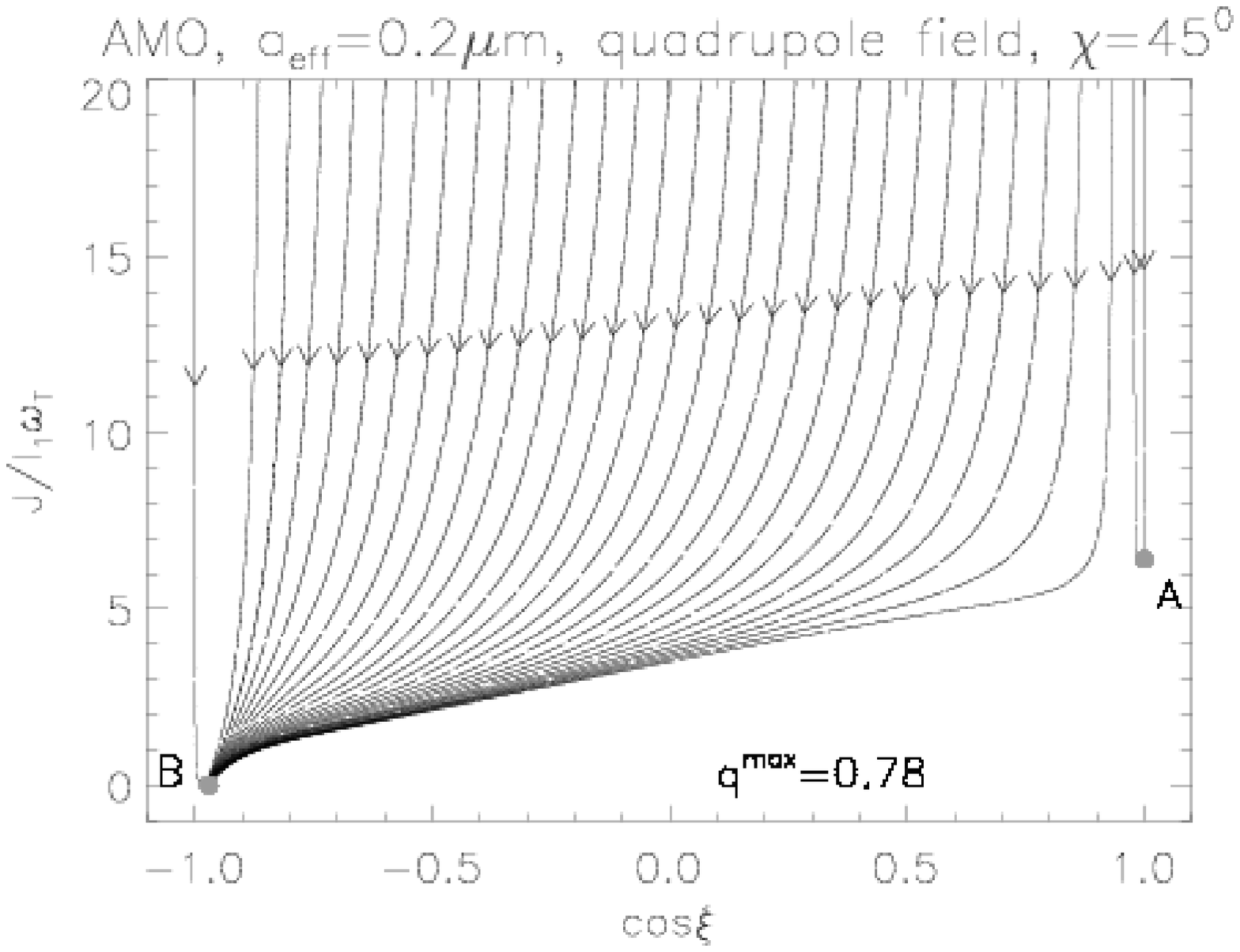}
\caption{RAT alignment by the quadrupole field with $q^{max}=0.78$ for $\chi=0^{\circ}$ ({\it upper panel}) and $45^{\circ}$ ({\it lower panel}), respectively. In the upper panel, A is a high-$J$ repellor, and B is a low-$J$ attractor, both corresponding to the perfect alignment of $\bJ$ with respect to $\bB$. In the lower panel, A becomes a high-J attractor and B is a low-J attractor.}
\label{f12}
\end{figure}

Let us consider now the RAT alignment by the quadrupole component of radiation field. The spatial distribution of the energy density for the quadrupole field is given by
\bea
u_{quad}=u_{ISRF}\ms^{2}\psi_{k}\mc^{2}\psi_{k},\label{eq14}
\ena
where $\psi_{k}$ is the angle between radiation direction and the symmetric axis of the quadrupole.
Following the same procedure as with the dipole field, we obtain results for the AMO with $q^{max}=0.78$ (i.e., similar to the value of the shape 2 at $\lambda=1.2 \mu m$, see LH07a) and the grain size $a_{eff}=0.2\mu m$ in Figure \ref{f12} for two directions of the quadrupole with the magnetic field: $\chi=0^{\circ}$ ({\it upper panel}) and $45^{\circ}$ ({\it lower panel}). The alignment occurs without high-$J$ attractors in the former case, but with a high-$J$ attractor in the later one.

\subsection{Simultaneous action of dipole and quadrupole components}

\begin{figure}
\includegraphics[width=0.5\textwidth]{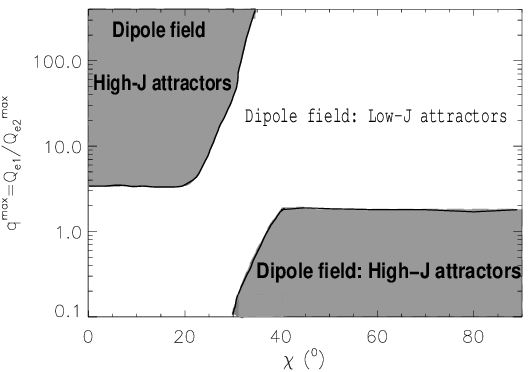}
\includegraphics[width=0.52\textwidth]{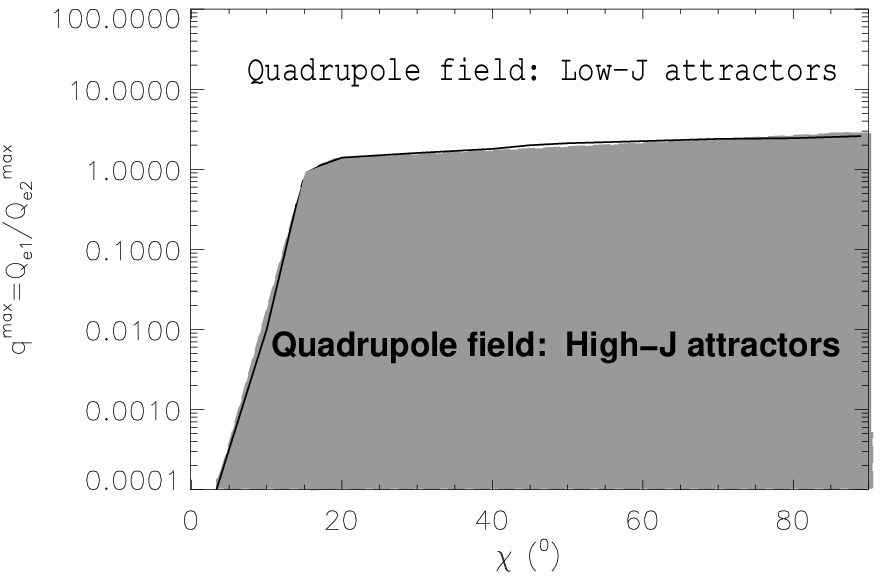}
\caption{Diagram depicting the existence of high-$J$ and low-$J$ attractors depending on $q^{max}$ and $\chi$ for dipole field ({\it Upper panel}) and quadrupole field ({\it Lower panel}). The shaded area represents the domain of the co-existence of high-$J$ attractors and low- $J$ attractors,  and the white area represents the presence of low-$J$ attractors only.}
\label{f13}
\end{figure}

Figure \ref{f13} shows the existence of high-$J$ and low-$J$ attractors as functions of $q^{max}$ and $\chi$ for the dipole ({\it upper panel})  and quadrupole ({\it lower panel})  fields. For the dipole field, it can be seen that the high-$J$ attractor appears when $q^{max}>3.2$ for $\chi=0^{\circ}$, then increases with $\chi$ increasing, and $q^{max}>100$ for $\chi=30^{\circ}$. For $\chi>30^{\circ}$, the high-$J$ attractor exists when $q^{max}<1.8$.  On the other hand, for the quadrupole field, there is the high-$J$ attractor when $q^{max}<2.5$ for a wide range  $\chi>20^{\circ}$. 

LH07a have calculated RATs for a number of irregular grain shapes, grain size and wavelength, and found that for the majority of shapes we have $q^{max} <10$. Thereby, from Figure \ref{f13} we see that the possibility of alignment with high-$J$ attractor is enhanced for both the dipole and quadrupole fields compared to a single beam. 

\begin{figure}
\includegraphics[width=0.5\textwidth]{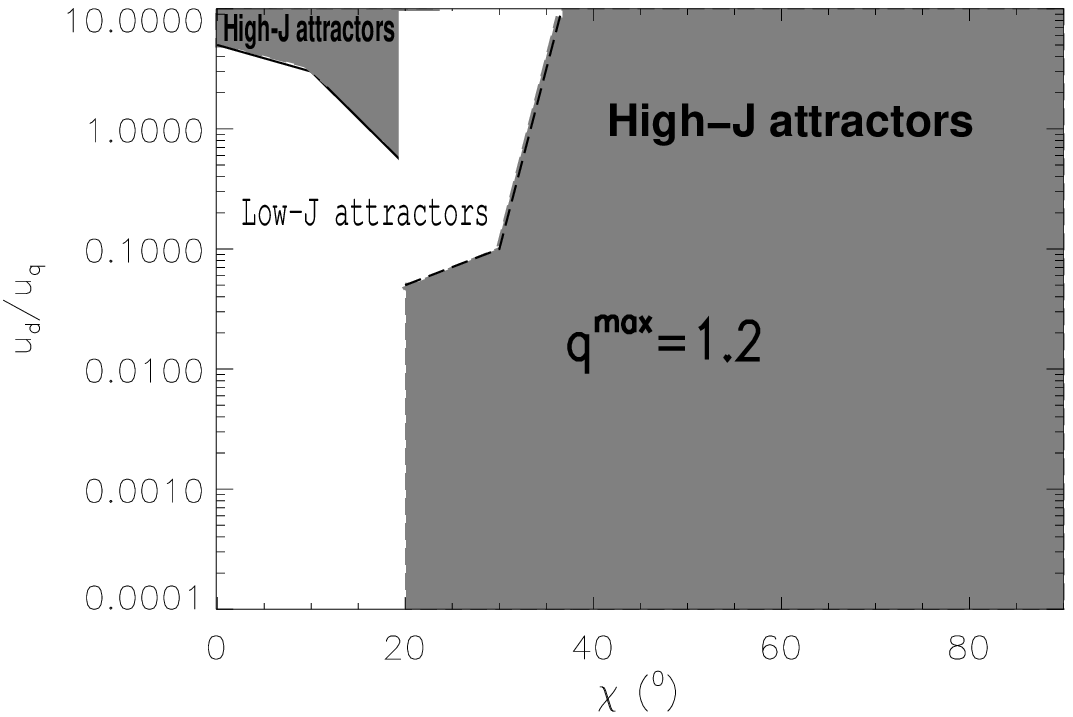}
\includegraphics[width=0.5\textwidth]{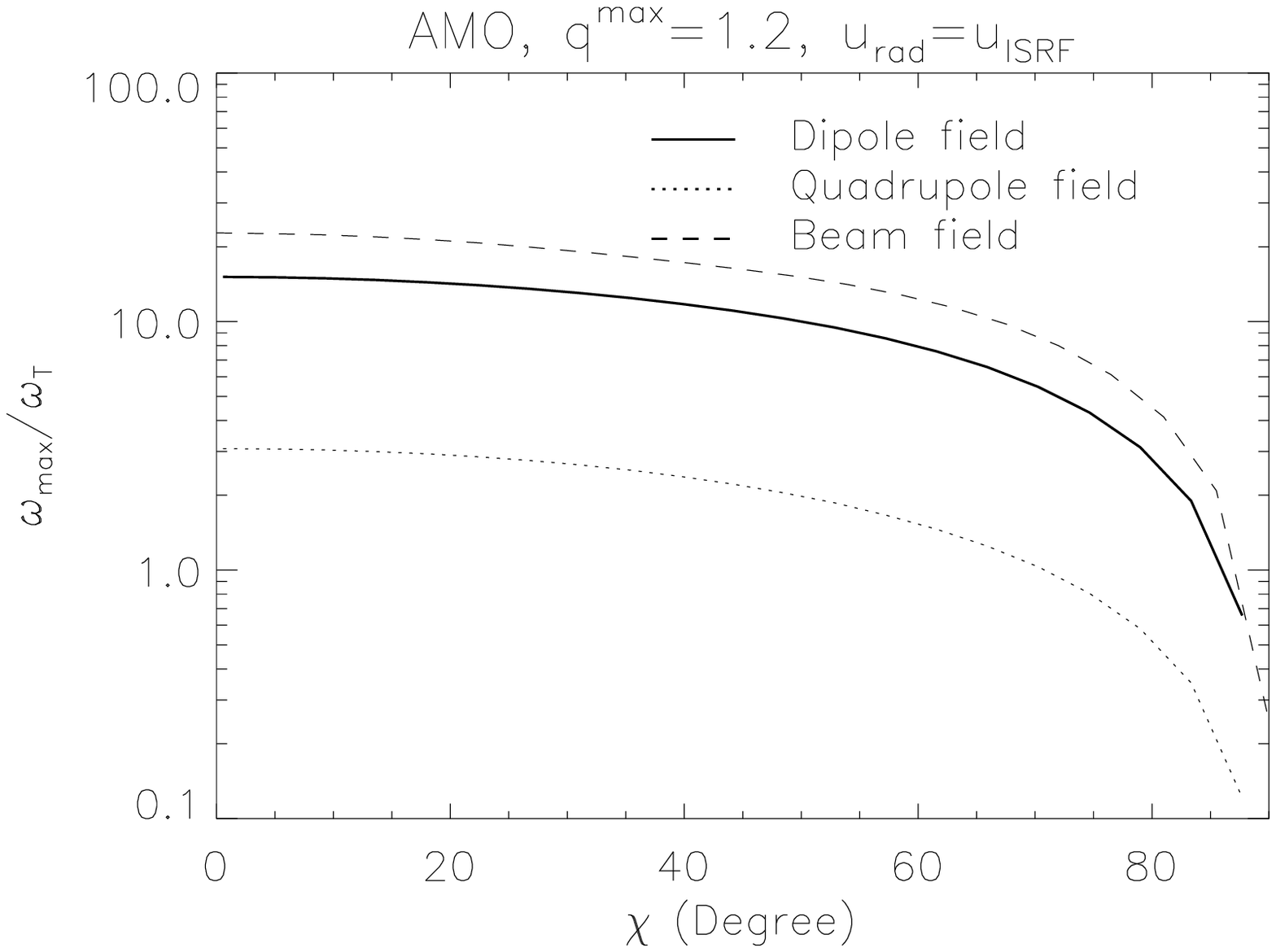}
\caption{{\it Upper panel:} Ratio of energy density of dipole to quadrupole components, $u_{d}/u_{q}$, for which grains are aligned with high-$J$ attractors for the AMO with $q^{max}=1.2$. The shaded area represents the domain of the co-existence of high-$J$ attractors and low -$J$ attractors, and the white area represents the presence of low-$J$ attractors only. {\it Lower panel:} The maximal value of angular velocity for a grain size $0.2\mu m$, as a function of $\chi$ for dipole and quadrupole fields and one beam field. For the later, $\chi$ denotes $\psi$ the angle between the beam direction and the magnetic field.}
\label{f14}
\end{figure}

In the upper panel of Figure \ref{f14}, we present the ratio of energy density of dipole to quadrupole components, $u_{d}/u_{q}$, as a function of $\chi$ for the AMO with $q^{max}=1.2$. For this AMO, the high-$J$ attractor appears frequently (see Fig. \ref{f13}{\it upper}). It reveals that the energy density of dipole component is required to be dominant over that of quadrupole in order to have high-$J$ attractor for $\chi<20^{\circ}$. In the range $\chi>30^{\circ}$, the high-$J$ attractor does not depend on the ratio of their energy density, because both components produce high-$J$ attractors.

Apart from the existence of high-$J$ attractors, the value of angular momentum $J$ at theses points is essential for calculations of degree of alignment. It is easy to see that the angular momentum achievable (i.e., maximal value) depends on the angle $\chi$.

Now, let us adopt the irregular grain (shape 1) with RATs obtained using DDSCAT, and calculate the value of the maximal angular momentum $J_{max}(\chi)=I_{1}\omega_{max}(\chi)$ produced by dipole and quadrupole components of the ISRF as a function of the angle $\chi$. The anisotropy degree of radiation $\gamma=0.1$ is assumed for all components. We comapre the results with that induced by a beam of radiation. 

In the lower panel of Figure \ref{f14} we present the resulting value of $\omega_{max}(\chi)$. There it can be seen that $\omega_{max}(\chi)$ decreases rapidly with increasing $\chi$ for the dipole component, but it exhibits slow change for the quadrupole component. In addition, $\omega_{max}$ induced by the dipole component is $\sim 10$ times greater than that by the quadrupole one. Therefore, when we only focus on the effect of spin-up by RATs, the quadrupole component can be neglected. The dipole component results in lower values of $\omega_{max}$ compared to that produced by the radiation beam (see Fig. \ref{f14}, lower), as a result of the average over the entire space for the dipole field.

\subsection{Implications for grain alignment}

The degree of RAT alignment depends strongly on the possibility of existence of high-$J$ attractors (see HL08a). In LH07a, we identified the criteria for the existence of high-$J$ as a function of the angle between the beam direction and the magnetic field. There it is shown that the high-$J$ attractor occurs when $q^{max}>2$ for $\psi<45^{\circ}$, and $q^{max}<1$ for $\psi>45^{\circ}$. In other words, there is a gap of $q^{max}$ from $1$ to $2$, in which there is no high-$J$ attractors, irrespective of grain shape and size. Fortunately, the dipole and quadrupole radiation can produce the high-$J$ attractors for $q^{max}<2$ (see Fig. \ref{f14}) for $\chi>30^{\circ}$. It indicates that the dipole and quadrupole extends the range of high-$J$ attractor and enable us to have high-$J$ attractors for $q^{max}=1$ to $2$. This range is easily satisfied for irregular grains (see LH07a) in a wide range of $\lambda/a_{eff}$. Therfore, an enhancement of the degree of the RAT alignment is expected for the dipole and quadrupole field compared to the case of a radiation beam.

\section{Discussion}
In the rest of this section we provide a brief account of our accomplishments in the paper and present our outlook on
the further work in the field of grain alignment. 

\subsection{Alignment in environments different from ISM: large grains}

Traditionally, grain alignment was the topic of interstellar medium (ISM) research. The gap between the studies of polarized radiation from aligned  dust in environments other then the ISM and the theory of grain alignment got so wide that the researchers outside the ISM domain sometimes write papers about aligned dust and do not refer to {\it any} theoretical work on grain alignment. However, there is ample evidence of grain alignment in non-ISM environments (see Tamura et al. 1999; Rosenbush et al. 2007; Hough et al. 2007). 

In many of these environments, e.g., for dust in comets and circumstellar dust, the radiation field is stronger than in the ISM, thus RATs are stronger. One difference that one faces there is that grains may be substantially larger. As we demonstrated
in the paper, for larger grains the internal relaxation gets slower, which makes one wonder whether the RAT alignment happens very differently from the ISM case. Our study shows that the direction of alignment of the angular momentum $\bJ$ with respect to the magnetic field $\bB$ is similar to that in the presence of strong internal relaxation. This alignment can be perfect with high-J and/or low-J attractors, depending on the factor $q^{max}$. However, the alignment of grain axes with respect to $\bJ$ depends on the initial angle between them, more precisely, on the initial value of dimensionless parameter $p$. If this angle is smaller than a particular value, $p_{4}$, defined in the main text, which is a function of $q^{max}$, the grain axis of maximum moment of inertia  $\ba_{1}$ gets aligned along $\bJ$, and both $\ba_{1}$ and $\bJ$ become aligned with $\bB$. HL08b showed that pinwheel torques (e.g., H$_{2}$ formation, isotropic radiative torque, etc.) increase fast with the grain size. As a result, large grains can be spun up to suprathermal rotation with $J\gg J_{th}$. The direct effect of suprathermal rotation is the alignment of $\ba_{1}$ and $\bJ$. 

Table 1 compares our results for grains without and with internal relaxation using the AMO. We can see that for high-J attractors, the alignment is the same in both cases. We note that the results in the later case were shown to be consistent with the alignment of irregular grains obtained by DDSCAT. However the general case of alignment of large irregular grains in the absence of internal relaxation requires further studies to confirm the predictions obtained with the AMO. Further studies should consider also the effect of pinwheel torques on the alignment in the absence of internal relaxation.

\subsection{Utility of AMO}

The analytical model (AMO) was introduced in LH07a to describe the case of alignment in the presence of {\it perfect} alignment of ${\bf J}$ with the axis of maximal moment of grain inertia. It provided excellent representation of the alignment of irregular grains within this approximation. Later in HL08a and HL08b we applied AMO to grains in the presence of thermal fluctuations, which at small values of $J$ partially randomize the alignment of ${\bf J}$ with the aforementioned axis. Nevertheless, AMO happened to do a nice job in these cases, adequately describing the behavior of irregular grains.

The case of no internal alignment is the extreme case of testing AMO. For instance, AMO has only one helicity axis, while multiple helicity axes are possible for an irregular grain. It is interesting, that even for this case, our limited testing is indicative of the AMO's utility.

\subsection{Towards modeling of polarized radiation}

Similar to the polarimetric work of non-ISM observers, the modeling of polarization arising from aligned grains has been mostly developing without much connection to the theory of grain alignment. Exceptions from this rule include
Cho \& Lazarian (2005, 2007), Pelkonen et al. (2007), Bethell et al. (2007), and Falceta-Golcaves et al. (2008). However, this modeling was done on the basis
of somewhat ad hoc alignment prescriptions, which should be improved as the theory gets predictive.

A step towards a more reliable polarization modeling is done in this paper where we considered grain alignment induced by
the dipole and quadroupole components\footnote{The alternative to modeling the RAT alignment by the multipoles of the radiation
field is to study the alignment numerically at every point of the data cube using the radiation field at the particular
point. Naturally, this would entail much more intensive computations.}  of the radiation field. The earlier studies assumed that the radiation was coming from a particular direction, which is, for instance, not the case for most of the grains in starless cores or accretion disks. In the latter cases it is proper to decompose the radiation field in multipoles and consider the effects of the individual components. Our study shows that the parameter space for having high-$J$ attractors differs for the alignment by the different multipole components.

Moreover we identified the range of torque ratio, $q^{max}$, for which the presence of dipole and quadroupole components of radiation field results in the alignment with high-$J$ attractors. The required range $q^{max}$ is fulfilled for irregular grains studied in HL08a.

\subsection{Credit to Dolginov \& Mytrophanov 1976}

In our paper we showed that the analytical results on RATs in Dolginov \& Mytrophanov (1976) were incorrect (see Fig. \ref{f15}). Therefore, naturally, our present results on the RAT alignment obtained for grains in the absence of internal relaxation, which was also the assumption in Dolginov \& Mytrophanov (1976) study, differ from those in the study.
This should not, however, undermine the pivotal significance of the the Dolginov \& Mytrophanov paper\footnote{Incidentally, 
the same paper discusses for
the first time the Barnett effect in the application to insterstellar grains. This induced the all-important notion of fast
Larmor precession of grains and helped E. Purcell to discover the effect of Barnett relaxation.}. This paper is
important as it {\it discovered} RATs and discussed the possibility of irregular grains being aligned by RATs, even if it failed to describe it quantitatively. In addition, the notion of grain helicity, which is the central concept of AMO, can be traced back to Dolginov \& Mytrophanov (1976) work. Moreover, as we mentioned in the introduction, the problem was so hard that the papers that followed Dolginov \& Mytraphanov (1976), in all its complexity, were not able to capture correctly the physics of the RAT alignment either.

\subsection{Other processes}

Our paper, for the sake of simplicity, does not consider the effect of the pinwheel torques. Such torques, e.g. torques
arising due to H$_2$ formation over catalytic sites over grain surface (Purcell 1979), were discussed in the framework of
the RAT alignment in Hoang \& Lazarian (2008b) for grains with strong internal relaxation. A study of effect of these torques
on large grains, for which the effects of internal relaxation are reduced, will be done elsewhere.
In addition, in our paper we considered the RAT alignment in respect to magnetic field. As we discussed in LH07a, for a
sufficiently slow rate of Larmor rotation, the alignment can happen also in respect to the radiation field. If, however,
grains have superparamagnetic inclusions the rate of rotation increases substantially. This, potentially, provides another way of testing whether grains have or do not have superparamagnetic inclusions.  

\section{Summary}

In this paper we continued our work on the RAT alignment using both AMO and DDSCAT calculations of torques.
Our principal results in the paper above can be summarized as follows:

$\bullet$ We identified the range of grain size for which the internal relaxation within a normal paramagnetic grain is not important. 

$\bullet$ We demonstrated that, in the absence of internal relaxation, RATs can align the grain's angular momentum with respect to the magnetic field, which is similar to the alignment in the presence of strong internal relaxation. For the internal alignment of grain axes with the angular momentum, it can be perfect, i.e. $\ba_{1}\|\bJ$ when the initial angle between them is small.

$\bullet $ We studied the RAT alignment induced by dipole and quadrupole components of the radiation field assuming perfect internal alignment of grain axis of maximum moment of inertia with the angular momentum due to {\it strong internal relaxation}. Using the AMO, we found that the parameter space for the existence of high-J attractors is extended  compared to the earlier studied case of a single direction radiation. This parameter space is given by the range of $q^{max}$ and the angle $\chi$ between the symmetric axis of dipole and quadrupole radiation fields and the magnetic field. Therefore, higher degrees of radiative alignment are expected.

$\bullet$ Our study for the joint action of dipole and quadrupole components showed that for the angle $\chi$ between the dipole and quadrupole axis and the radiation direction smaller than $\sim 20$ degree, the dipole component has to be dominant over quadrupole one in order to align grains with high-J attractors.

\section*{Acknowledgments}
We acknowledge the support by the NSF Center for Magnetic Self-Organization in Laboratory and Astrophysical Plasmas and NSF grant AST 0507164.

\appendix
\clearpage
\begin{table}
\caption{Comparison of radiative alignment for AMO}
\begin{tabular}{llcll} \hline\hline\\
\multicolumn{1}{c}{\it Without internal relaxation (this work)} & & \multicolumn{1}{c}{\it With internal relaxation (LH07a)}\\[1mm]

\hline\\
{{\bf High-J attractors}}& {{\bf Low-J attractors}} &{{\bf High-J attractors}}&{{\bf Low-J attractors}}\\[1mm]
&&&&\\[1mm]
{$\bJ\|\bB$}& {$\bJ$ aligned parallel or} &{$\bJ\| \bB$}&{$\bJ$ aligned parallel or at}\\[1mm]
&{at some angle with $\bB$}&&{at some angle with $\bB$}\\[1mm]
{Long axes $\perp \bB$}& {Long axes $\perp$ or $\|$ $\bJ$} &{Long axes $\perp \bB$}&{Long axes $\perp \bB$}\\[1mm]\\[1mm]
\hline\hline\\
\end{tabular}

\end{table}
\section{RATs for the analytical model: AMO}
RAT for the toy model in Figure \ref{f1} is given by
\bea
{\bf \Gamma}_{rad}&=\frac{\gamma u_{rad}\lambda l_{2}^{2}}{2}{\bf Q}_{\Gamma},\label{a1}
\ena
where ${\bf Q}_{\Gamma}=(Q_{e_{1}}{\bf e}_{1}+Q_{e_{2}}{\bf e}_{2}+Q_{e_{3}}{\bf e}_{3})$ is the vector of RAT efficiency, with $Q_{e_{1}}, Q_{e_{2}}$ and $Q_{e_{3}}$ the components of ${\bf Q}_{\Gamma}$ in the laboratory system. Here $l_{2}$ is the size of the squared mirror, $\lambda$ is the wavelength, and $u_{rad}$ is the energy density in unit erg cm$^{-3}$ of the radiation field.

Using the self-similar scaling of the magnitude of RATs obtained for an irregular grain of size $a$ illuminated by radiation field of wavelength $\lambda$,
\bea
\left|Q_{\Gamma}\right|&\sim& 0.4\left(\frac{\lambda}{a}\right)^{-3} \mbox{ for~$\lambda > 1.8 a$},\label{a2}\\
&\sim& 0.4 \mbox { for~$\lambda \le 1.8 a$},\label{a3}
\ena
and the functional forms of RATs from the AMO, we can write RAT components as following
\bea
Q_{e_{1}}(\Theta, \beta,\Phi= 0)&=&\frac{\left|Q_{\Gamma}\right|q^{max}}{\sqrt{(q^{max})^{2}+1}}\frac{q_{e_{1}}(\Theta, \beta, \Phi=0)}{q_{e_{1}}^{max}},\label{a4}\\
Q_{e_{2}}(\Theta, \beta, \Phi=0)&=&\frac{\left|Q_{\Gamma}\right|}{\sqrt{(q^{max})^{2}+1}}\frac{q_{e_{2}}(\Theta, \beta, \Phi=0)}{q_{e_{2}}^{max}},\label{a5}\\
Q_{e_{3}}(\Theta, \beta, \Phi=0)&=&\frac{\left|Q_{\Gamma}\right|q^{max}}{\sqrt{(q^{max})^{2}+1}}\frac{q_{e_{3}}(\Theta, \beta, \Phi=0)}{q_{e_{3}}^{max}},\label{a6}
\ena
where
\bea
q_{e_{1}}(\Theta, \beta, \Phi=0)&=&-\frac{4l_{1}}{\lambda}C\left(n_{1}n_{2}\mc^{2}\Theta+\frac{n_{1}^{2}}{2}\mc\beta\ms2\Theta 
-\frac{n_{2}^{2}}{2}\mc\beta\ms2\Theta-n_{1}n_{2}\ms^{2}\Theta\mc^{2}\beta\right),\label{ab7}\\
q_{e_{2}}(\Theta, \beta,\Phi= 0)&=&\frac{4l_{1}}{\lambda}C\left(n_{1}^{2}\mc\beta\mc^{2}\Theta-\frac{n_{1}n_{2}}{2}\mc^{2}\beta\ms2\Theta-\frac{n_{1}n_{2}}{2}\ms2\Theta +n_{2}^{2}\mc\beta\ms^{2}\Theta\right),\label{ab8}\\
q_{e_{3}}(\Theta, \beta,\Phi= 0)&=&\frac{4l_{1}}{\lambda}C n_{1}\ms\beta \left[n_{1}\mc\Theta-n_{2}\mc\beta\ms\Theta\right]+\left(\frac{b}{l_{2}}\right)^{2}\frac{2e a}{\lambda}(s^{2}-1)K(\Theta)\ms 2\Theta,\label{ab9}
\ena
with $q_{e_{j}}^{max}=max\langle q_{e_{j}}(\Theta, \beta,\Phi= 0)\rangle_{\beta}$ for $j=1,2$ and $3$. The ratio of torque components is then defined by
\bea
q^{max}=\frac{max{\langle Q_{e_{1}}(\Theta, \beta, \Phi=0)\rangle_{\beta}}}{max{\langle Q_{e_{2}}(\Theta, \beta, \Phi=0)\rangle_{\beta}}}.\label{qfact}
\ena
In equations (\ref{ab7})-(\ref{ab9}), $C$ is a function defined as
\bea
C=\left|n_{1}\mc\Theta-n_{2}\ms\Theta\mc\beta\right|,
\ena
where  $\Theta$ is the angle between the axis of maximum moment of inertia ${\bf a}_{1}$ and the radiation direction ${\bf k}$, $\beta$ is the angle describing the rotation of the grain about $\ba_{1}$ (see Fig. \ref{f2}{\it lower}); $n_{1}=-\ms \alpha, n_{2}=\mc \alpha$ are components of the normal vector
of the mirror tilted by an angle $\alpha$ in the grain coordinate system, $a, b$ are minor and major semi-axes of the spheroid, $s=a/b<1$ and $e$ is the eccentricity of the spheroid, $l_{1}$ is the distance from the mirror to the spheroid, and $l_{2}$ is the size of the squared mirror; $K(\Theta)$ is the fitting function (see also LH07a). The second term of equation (\ref{ab9}) represents the torque due to the spheroid. Assuming $l_{1}\sim \lambda$ and $b, a \sim l_{2} \ll l_{1}$, then this term is subdominant compared to the first term (see Fig. \ref{f1}). Thus, we disregard it in our calculations.

Our calculations for the alignment in the presence of thermal fluctuations  showed that the AMO can reproduce the alignment property with low-J as found with RATs obtained by DDSCAT when $q_{e_{1}}$ is modified to (see HL08a)
\bea
q_{e_{1}}(\Theta, \beta, \Phi=0)=-\frac{4l_{1}}{\lambda}C\left(n_{1}n_{2}\frac{[3\mc^{2}\Theta-1]}{2}+\frac{n_{1}^{2}}{2}\mc\beta\ms2\Theta 
-\frac{n_{2}^{2}}{2}\mc\beta\ms2\Theta-{n_{1}n_{2}}\mc2\beta\right).\label{a10}
\ena
This modification can arise from the imperfect scattering and/or the absorption effect by the mirror (LH07a). Also, the results in LH07a remain unchanged because the averaging over $\beta$ for the last term goes to zero.

We adopt the AMO with $\alpha=45^{\circ}$ in this paper, unless mentioned otherwise. We also assume the amplitude of $Q_{e_{3}}$ is comparable to that of $Q_{e_{1}}$ and $Q_{e_{2}}$.

RATs at a precession angle $\Phi$ (see Fig. \ref{f2}{\it lower}) can be derived from RATs at $\Phi=0$ using the coordinate system transformation, as follows:
\bea
Q_{e_{1}}(\Theta, \beta, \Phi)&=&Q_{e_{1}}(\Theta, \beta, \Phi=0),\label{aeq4}\\ 
Q_{e_{2}}(\Theta, \beta, \Phi)&=&Q_{e_{2}}(\Theta, \beta,\Phi=
0)\mbox{cos}\Phi+Q_{e_{3}}(\Theta, \beta,\Phi=
0)\mbox{sin}\Phi,\label{aeq5} 
\\  
Q_{e_{3}}(\Theta, \beta, \Phi)&=&Q_{e2}(\Theta, \beta,\Phi=
0)\mbox{sin}\Phi-Q_{e_{3}}(\Theta, \beta,\Phi=
0)\mbox{cos}\Phi.\label{aeq6}  
\ena

\subsection{AMO versus  the model of Dolginov \& Mytrophanov 1976}
\begin{figure}
\includegraphics[width=0.49\textwidth]{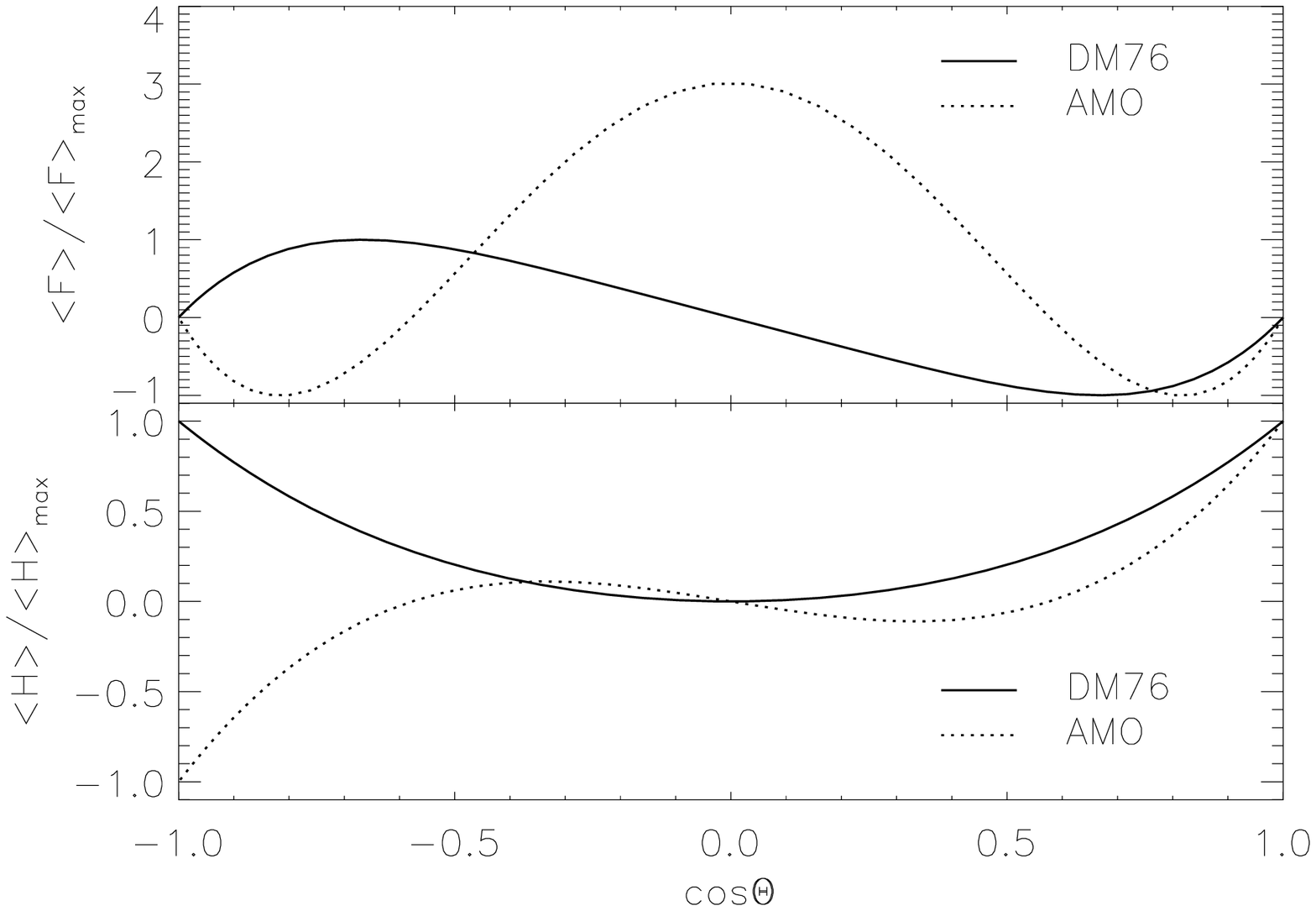}
\includegraphics[width=0.49\textwidth]{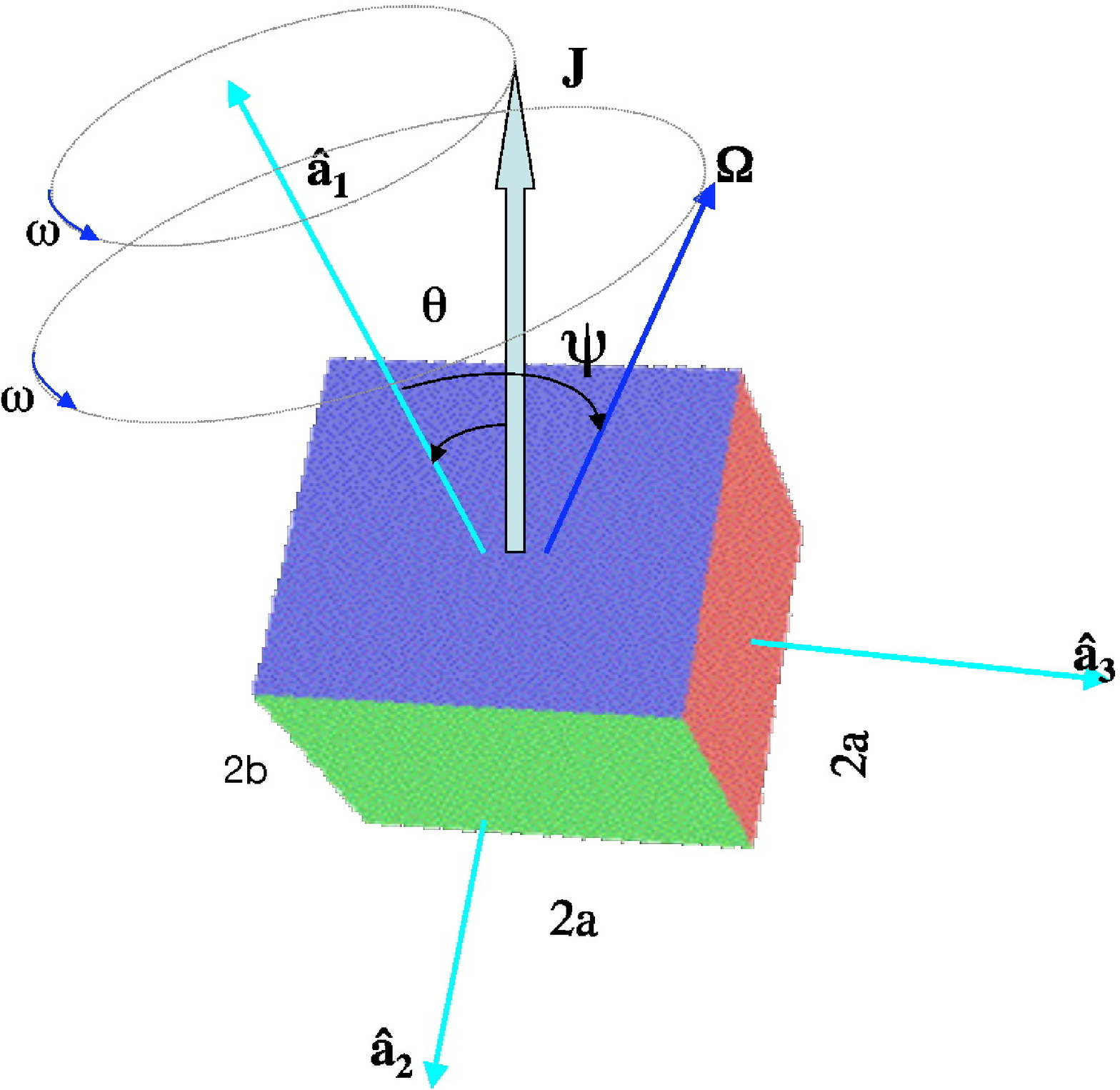}
\caption{{\it Upper panel:}Comparison of RATs components for AMO and DM76. $\Theta$ is the angle between the axis of maximum moment of inertia (shortest) with the radiation in AMO, and is the angle between the twisted axis (longest axis)  and radiation for DM76. Stationary points $\Theta=0, \pi$ appear in both models, but the form of torques are different. $\langle F\rangle$ is symmetric for the AMO, but not for SM76, also in DM76, $\langle H \rangle$ is symmetric instead.
{\it Lower panel:}A brick grain adopted in calculations for internal relaxation timescale.} 
\label{f15}
\end{figure}

Figure \ref{f15} shows the comparison of the RAT components for the AMO and RATs obtained for the twisted spheroid from Dolginov \& Mytrophanov (1976). It can be seen that the torques are radically different. According to LH07a, the AMO corresponds to calculations of torques for irregular grains, thus we can conclude that the model in Dolginov \& Mytrophanov (1976) does not represent adequately RATs.

\section{Dynamical equations for irregular grains}
For irregular grains, we define a parameter which is constant during the free-torque motion:
\bea
p=\frac{2I_{1}E}{J^{2}},\label{a6}
\ena
where $E$ is the total energy, $I_{1}$ is the inertia moment along the principal axis $\ba_{1}$ and $J$ is the value of angular momentum. Since both $J$ and $E$ are conserved during the torque-free motion, $p$ is accordingly conserved.
The evolution of $p$ in time due to external torques is then
\bea
\frac{dp}{dt}=\frac{1}{J^{2}}\left(2I_{1}\frac{dE}{dt}J-2I_{1}E\frac{dJ}{dt}\right)\label{a7}
\ena
The energy of the grains varies as
\bea
\frac{dE}{dt}=\left({\bf J}.\frac{d\omega}{dt}+\Gamma.\omega\right)\label{a8}
\ena
The first term describes the energy dissipation due to the Barnett and nuclear relaxation. The second terms represents the effect of external torques on the grain rotational energy. This equation differs from that in WD03 by a factor of 2. We consider here only RATs, so $\Gamma=\Gamma_{rad}=M Q_{\Gamma}$.

Substituting $dE/dt$ into equation (\ref{a7}), we get

\bea
\frac{dp}{dt}=\frac{1}{J^{2}}\left[I_{1}\left({\bf J}.\frac{d\omega}{dt}+\Gamma.\omega\right) -p {\bf J}.\Gamma\right].
\label{a9}
\ena
Averaging equation (\ref{a9}) over the torque-free motion, we obtain 
\bea
\frac{J^{2} dp}{dt}=MI_{1}\langle Q_{\omega}\rangle-pJ\frac{dJ}{dt}-\frac{(p-1)}{t_{int}}\left(\frac{1-pI_{3}/I_{1}}{1-I_{3}/I_{1}}\right),\label{a10}
\ena
where we used ${\bf \Gamma}=d{\bf J}/dt$, and 
\bea
Q_{\omega}={\bf Q}_{\Gamma}.{\bf \omega}=Q_{a_{1}}\omega_{1}+Q_{a_{2}}\omega_{2}+Q_{a_{3}}\omega_{3}.\label{qome}
\ena
Here $\omega_{i}$ for i=1,2 and 3 corresponds to the components of angular velocity along three grain axes $\ba_{1}, \ba_{2}$ and $\ba_{3}$.

\section{RAT alignment induced by a single component for the spheroidal AMO}
We study first the alignment with high-$J$ attractor. For this purpose, we consider the alignment by the first component $Q_{e_{1}}$, which is shown to produce the alignment  with high-$J$ attractor (see LH07a). With this simplification, we can obtain the analytical results for the motion.
According to equation (31) in HL08a, the averaged value of $Q_{e_{1}}$ over torque-free motion, and fast Larmor precession is given by
\bea
\langle Q_{e_{1}}\rangle_{\phi}&=&Q_{e_{1}}^{max}\left(3 \mc^{2}\xi\mc^{2}\theta+\frac{3 \ms^{2}\xi\ms^{2}\theta}{2}-1\right),\label{a11}\\
\langle Q_{a_{1}}\rangle_{\phi}&=&Q_{e_{1}}^{max}\mc\theta\mc\xi \left(-1+3 \mc^{2}\xi\mc^{2}\theta+\frac{9}{2}\ms^{2}\xi\ms^{2}\theta\right).\label{a12}
\ena
The aligning and spin-up torque components are then
\bea
\langle F\rangle_{\phi}=-\langle Q_{e_{1}}\rangle_{\phi}\ms\xi,\label{a13}\\
\langle H\rangle_{\phi}=\langle Q_{e_{1}}\rangle_{\phi}\mc\xi.\label{a14}
\ena
It can be seen that the torque components are functions of two alignment angles $\theta$ and $\xi$.

Let us investigate the property of these stationary points. Consider first the stationary points $\xi=\pi$ and $\theta=90^{\circ}$. Equation (\ref{a14}) shows that $\langle H\rangle_{\phi} =\langle Q_{e_{1}}^{max}\rangle^{2}>0$ because $\langle Q_{e_{1}}\rangle=-Q_{e_{1}}^{max}$. In addition, the first derivative of $\langle F\rangle_{\phi}<0$ (see eq. \ref{a13}). Therefore, $\xi=\pi$ is a high-$J$ attractor with $\theta=90^{\circ}$. Moreover, the stationary point $\xi=0, \theta=0$ is still a high-$J$ attractor because $\langle F'\rangle_{\phi}/\langle H\rangle_{\phi}<0$. This indicates that in the absence of internal dissipation, the alignment occurs in two types: the longest axis parallel and perpendicular to the magnetic field. The former is consistent with the Davis-Greenstein prediction, i.e., ``right'' alignment, while the later provides ``wrong'' alignment (see LH07a).

\begin{figure}
\includegraphics[width=0.49\textwidth]{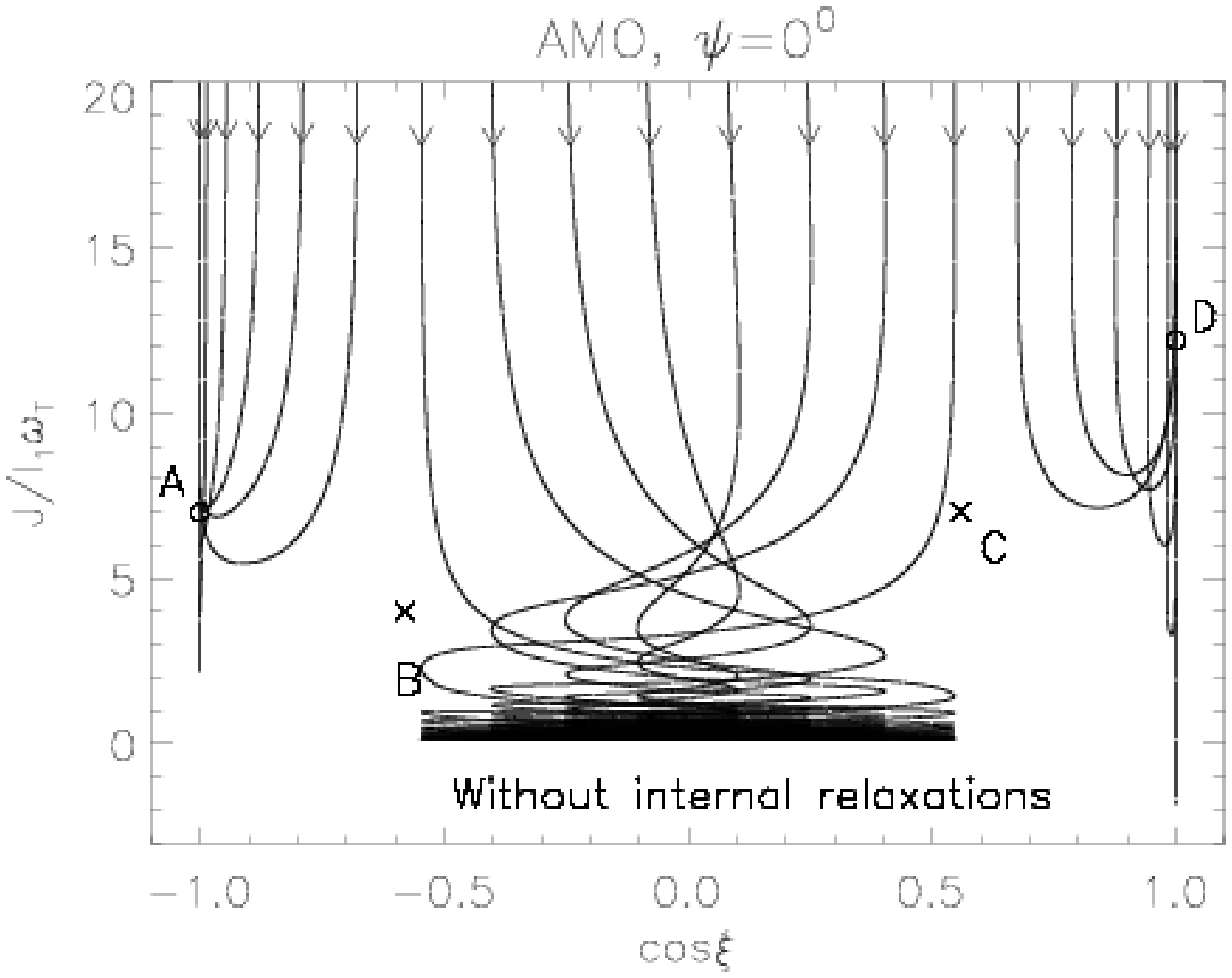}
\includegraphics[width=0.49\textwidth]{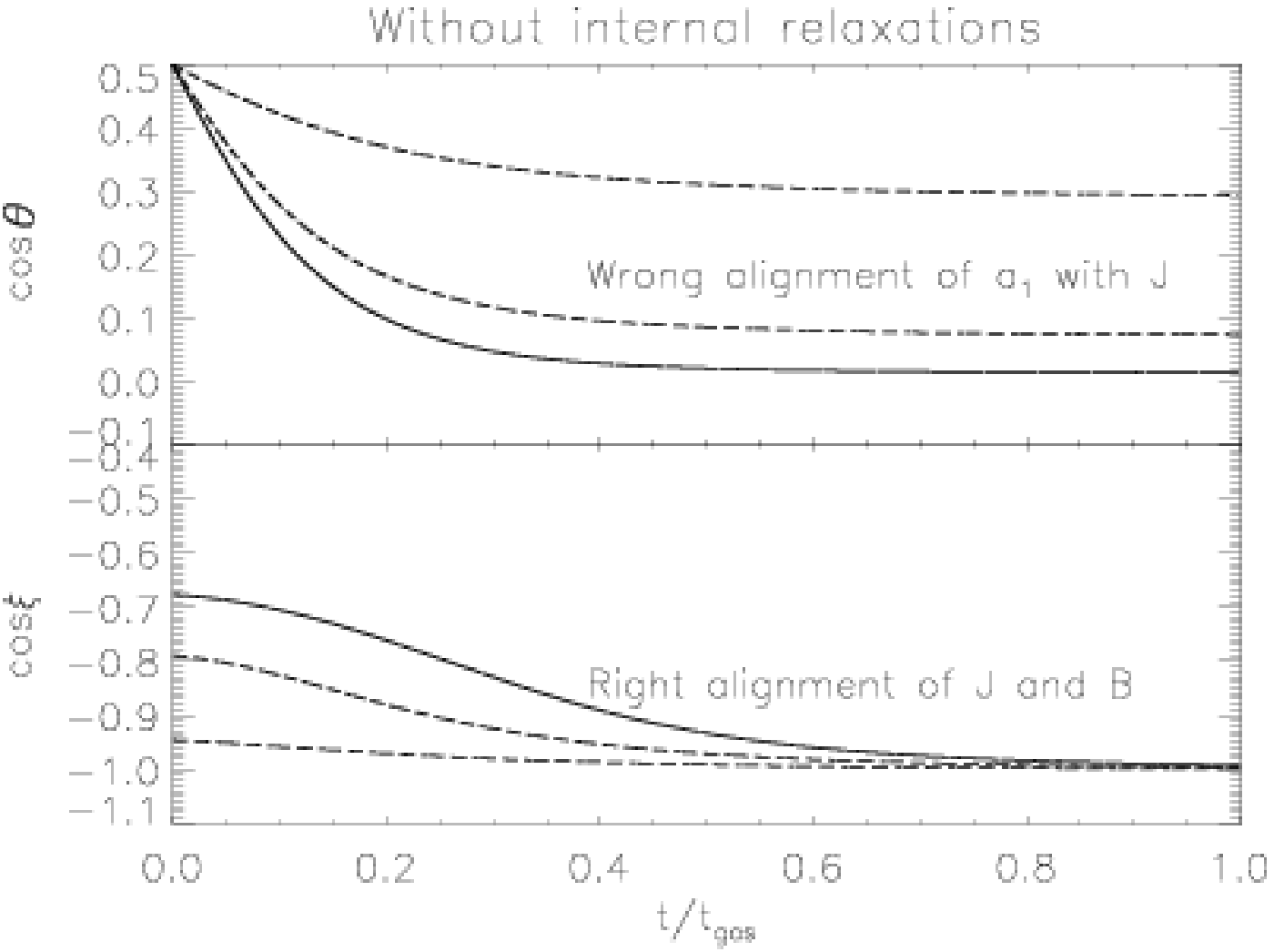}
\caption{{\it Upper panel: }RAT alignment by $Q_{e1}$, in the absence of internal relaxation. A and D denoted by circles are high-$J$ attractors, and C and D denoted by crosses are repellors. The alignment at A corresponds to $\ba_{1}\perp {\bf B}$, and $\ba_{1}\| {\bf B}$ for the alignment at D. {\it Lower panel:} Evolution of $\theta$ and $\xi$ for a few grains with $\mc\xi_{0}<-0.8$ as functions of time.}
\label{f16}
\end{figure}
The trajectory map in Figure \ref{f16} for the alignment by one component $Q_{e1}$ exhibits two high-$J$ attractors A and D, as expected. In addition, there are two repellors B and C. The final state of grains depends on their initial angles $\xi_{0}$. For instance, grains with initial angles $|\mc\xi_{0}|>0.6$ are aligned on A and D, but grains with $|\mc\xi_{0}|<0.6$ are constrained within two repellors B and C, and finally damped by gas friction.

The lower panel represents the evolution of $\theta$ and $\xi$ as functions of time for few grains with $\mc\xi_{0}<-0.8$. It can be seen that the angular momentum of these grains are perfectly aligned with respect to the magnetic field, but the shortest axis $\ba_{1}$ is nearly perpendicular to $\bJ$, corresponding to ``wrong'' internal alignment.

\section{Transformation of coordinate system}
To find the torques in the lab system ($\be_{i}$) when the radiation direction ${\bf k}$ varies, we need to implement the coordinate transformation from the system $k k_{x} k_{y}$ to $\be_{i}$ (see Fig. \ref{f10} ). We assume that the magnetic field is directed along $\be_{1}$, then the ${\bf k}_{z}=\bk$, and ${\bf k}_{y}$ lies in the plane $\bB, \bk$. Thus, $\bk_{x}$ is perpendicular to the plane $\bk,\bk_{y}$. Therefore, the coordinate system $\bk,\bk_{x},\bk_{y}$ acts as the system $\be_{i}$ in Figure \ref{f2}.

Denote $Q_{k}$ be torques components in the k-system, we need to know the torque components in the lab system. The transformation from ${\bk}$-system to ${\be_{i}}$-system is carried out by three rotations with three Euler angles.

Following Figure \ref{f10}, we have
\bea
\be_{1}^{\chi}=\be_{1}\mc\chi+\be_{2}\ms\chi,\label{a15}\\
\be_{1}^{\chi}=-\be_{1}\ms\chi+\be_{2}\mc\chi,\label{a16}\\
\be_{3}^{\chi}=\be_{3}\label{a17}
\ena

The radiation direction $\bk$ is determined by two angles $\psi_{k}$ and $\phi_{k}$ in the lab system, we have the coordinate transformation from $\bk$-system to ${\bf d}$-system: 
\bea
\be_{1}^{0}=\be_{1}^{\chi}\mc\psi_{k}+\be_{2}^{\chi}\ms\psi_{k}\mc\phi_{k}+\be_{3}^{\chi}\ms\psi_{k}\ms\phi_{k}\label{a18}\\
\be_{2}^{0}=-\be_{1}^{\chi}\ms\psi_{k}+\be_{2}^{\chi}\mc\psi_{k}\mc\phi_{k}+\be_{3}^{\chi}\mc\psi_{k}\ms\phi_{k}\label{a19}\\
\be_{3}^{0}=\be_{2}^{\chi}\ms\phi_{k}+\be_{3}^{\chi}\mc\phi_{k}^{\chi}\label{a20}
\ena

Assuming that $\ba_{1}$ is parallel to $\bJ$, which is described by $\xi, \phi$ in the lab system, then we have
\bea
\ba_{1}=\be_{1}\mc\xi+\be_{2}\ms\xi\mc\phi+\be_{3}\ms\xi\ms\phi\label{a21}\\
\ba_{2}=-\be_{1}\ms\xi+\be_{2}\mc\xi\mc\phi+\be_{3}\mc\xi\ms\phi,\label{a22}\\
\ena

We can find the angles $\Theta, \Phi $ and $\beta$ in the $\be_{i}^{0}$ system:
\bea
\mc\Theta=\be_{1}^{0}.\ba_{1},\label{a23}\\
tan\frac{\Phi}{2}=\frac{\ms\Theta-\ba_{1}.\be_{2}^{0}}{\ba_{1}.\be_{3}^{0}}\label{a24}\\
\beta=2{\mbox tan}^{-1} \left(\frac{\ms\Theta+\ba_{2}.\be_{2}^{0}}{\ms\Theta(\ba_{2}.\be_{3}^{0}\mc\Phi-\ba_{2}.\be_{2}^{0}\ms\Phi)}\right)\label{a25},
\ena
By substituting equations (\ref{a18})-(\ref{a20}) and (\ref{a21}) and (\ref{a22}) into (\ref{a23}) and (\ref{a25}), we obtain the angles $\Theta, \Phi$ and $\beta$, the torques $Q_{ei}^{0}$ are interpolated. Finally, we find the corresponding torques, $Q_{ei}(\xi, \phi)$ by coordinate transformation:
\bea
Q_{e_{i}}=C^{T}_{ij}B^{T}_{jk}Q_{e_{k}}^{0},\label{a26}
\ena
Here matrices $C^{T}, B^{T}$  are tranposal matrices of C and B, inferred from the equations (\ref{a23})-(\ref{a25}) and (\ref{a4})-(\ref{a6}) following
\bea
\be_{i}^{\chi}=C_{ij}\be_{j},\label{a27}\\
\be_{i}^{0}=B_{ij}\be_{j}^{\chi}.\label{a28}
\ena

\end{document}